\documentclass[manuscript]{aastex}

\shorttitle{Spectral Indices for MARVELS}
\shortauthors{Ghezzi et al.}

\begin{document}

\def\teff{$T_{\rm eff}$}
\def\logg{$\log{g}$}


\title{Accurate Atmospheric Parameters at Moderate Resolution Using
       Spectral Indices: Preliminary Application to the MARVELS Survey
       \footnote{Based on
       observations obtained with the 2.2 m MPG telescope 
       at the European Southern Observatory (La Silla,
       Chile), under the agreement ESO-Observat\'orio 
       Nacional/MCT, and the Sloan Digital Sky Survey,
       which is owned and operated by the Astrophysical 
       Research Consortium.}}


\author{Luan Ghezzi\altaffilmark{1,2}, 
        Let\'icia Dutra-Ferreira\altaffilmark{3,2},
        Diego Lorenzo-Oliveira\altaffilmark{3,2},
        Gustavo F. Porto de Mello\altaffilmark{3,2}, 
        Bas\'ilio X. Santiago\altaffilmark{4,2},
        Nathan De Lee\altaffilmark{5,6,7},
        Brian L. Lee\altaffilmark{7,8},
        Luiz N. da Costa\altaffilmark{1,2}, 
        Marcio A. G. Maia\altaffilmark{1,2},
        Ricardo L. C. Ogando\altaffilmark{1,2},
        John P. Wisniewski\altaffilmark{9},
        Jonay I. Gonz\'alez Hern\'andez\altaffilmark{10,11},
				Keivan G. Stassun\altaffilmark{5,6},
				Scott W. Fleming\altaffilmark{12},
				Donald P. Schneider\altaffilmark{13,14},
				Suvrath Mahadevan\altaffilmark{13,14},
				Phillip Cargile\altaffilmark{5},
        Jian Ge\altaffilmark{7},        
        Joshua Pepper\altaffilmark{15,5},
				\& Ji Wang\altaffilmark{16}
        }


\altaffiltext{1}{Observat\'orio Nacional, Rua Gal. Jos\'e Cristino 77, 
Rio de Janeiro, RJ 20921-400, Brazil}

\altaffiltext{2}{Laborat\'orio Interinstitucional de e-Astronomia - LIneA, 
Rua Gal. Jos\'e Cristino 77, Rio de Janeiro, RJ 20921-400, Brazil; luan@linea.gov.br}

\altaffiltext{3}{Universidade Federal do Rio de Janeiro, Observat\'orio do Valongo, 
Ladeira do Pedro Ant\^onio 43, Rio de Janeiro, RJ 20080-090, Brazil}

\altaffiltext{4}{Instituto de F\'isica, UFRGS, Caixa Postal 15051, Porto Alegre, 
RS 91501-970, Brazil}

\altaffiltext{5}{Department of Physics and Astronomy, Vanderbilt
University, Nashville, TN 37235, USA}

\altaffiltext{6}{Department of Physics, Fisk University, Nashville, TN, USA}

\altaffiltext{7}{Department of Astronomy, University of Florida, 211
Bryant Space Science Center, Gainesville, FL, 32611-2055, USA}

\altaffiltext{8}{Astronomy Department, University of Washington, Box
351580, Seattle, WA 98195, USA}

\altaffiltext{9}{H. L. Dodge Department of Physics and Astronomy,
University of Oklahoma, 440 West Brooks St Norman, OK 73019, USA}

\altaffiltext{10}{Instituto de Astrof\'isica de Canarias (IAC),
E-38205 La Laguna, Tenerife, Spain}

\altaffiltext{11}{Departamento de Astrof\'isica, Universidad de La
Laguna, 38206 La Laguna, Tenerife, Spain}

\altaffiltext{12}{Space Telescope Science Institute - STScI, 3700 San 
Martin Dr., Baltimore, MD 21218}

\altaffiltext{13}{Department of Astronomy and Astrophysics, The
Pennsylvania State University, University Park, PA 16802, USA}

\altaffiltext{14}{Center for Exoplanets and Habitable Worlds,
The Pennsylvania State University, University Park, PA 16802, USA}

\altaffiltext{15}{Department of Physics, Lehigh University, 16 Memorial 
Drive East, Bethlehem, PA 18015, USA}

\altaffiltext{16}{Department of Astronomy, Yale University, New Haven, 
CT 06511 USA}


\begin{abstract}

Studies of Galactic chemical and dynamical evolution in the solar neighborhood 
depend on the availability of precise atmospheric parameters (effective 
temperature \teff, metallicity [Fe/H] and surface gravity \logg) for solar-type stars.
Many large-scale spectroscopic surveys operate at low to moderate spectral resolution 
for efficiency in observing large samples, which makes the stellar characterization 
difficult due to the high degree of blending of spectral features. Most surveys therefore 
employ spectral synthesis, which is a powerful technique, but relies heavily on the completeness 
and accuracy of atomic line databases and can yield possibly correlated atmospheric parameters.
In this work, we use an alternative method
based on spectral indices to determine the atmospheric parameters of a sample of 
nearby FGK dwarfs and subgiants observed by the MARVELS survey 
at moderate resolving power (R $\sim$ 12,000). To avoid a 
time-consuming manual analysis, we have developed three codes to 
automatically normalize the observed spectra, measure the equivalent widths of 
the indices and, through the comparison of those with values calculated with pre-determined 
calibrations, determine the atmospheric parameters of the stars. The calibrations were 
derived using a sample of 309 stars with precise stellar parameters 
obtained from the analysis of high-resolution FEROS spectra, 
permitting the low-resolution equivalent widths to be directly related to the stellar parameters.
A validation test of the method was conducted with a sample of 30 MARVELS targets that 
also have reliable atmospheric parameters derived from the high-resolution spectra and 
spectroscopic analysis based on excitation and ionization equilibria method.
Our approach was able to recover the parameters within 80 K for \teff, 0.05 dex for [Fe/H] and 
0.15 dex for \logg, values that are lower or equal to the typical external uncertainties 
found between different high-resolution analyzes. 
An additional test was performed with a subsample of 138 stars from the ELODIE stellar library 
and the literature atmospheric parameters were recovered within 125 K for \teff, 0.10 dex for 
[Fe/H] and 0.29 dex for \logg. These precisions are consistent or better than those provided
by the pipelines of surveys operating with similar resolutions.
These results show that the spectral indices are a competitive tool to characterize stars with 
the intermediate resolution spectra.

\end{abstract}


\keywords{Stars: atmospheres --- Stars: fundamental parameters --- Stars: solar-type ---
          techniques: spectroscopic --- (Galaxy:) solar neighborhood}


\section{Introduction}

\label{intro}

Solar-type stars are prime targets for many studies in astrophysics.
Their spectra are rich in metallic lines, which allow precise determinations
of the fundamental stellar parameters and elemental abundances through 
different well-established techniques (e.g., excitation and ionization equilibria 
or spectral synthesis). They are also long-lived and have a large age 
dispersion, probing a considerable fraction of the history of the Milky Way. 
Furthermore, the compositions of their atmospheres remain almost unchanged 
(with the exception of Li, Be and B) throughout their evolution on the main 
sequence. All these properties make solar-type stars the ideal candidates 
to study time-dependent processes, such as the chemical evolution in the 
solar neighborhood, and many successful examples of this application can be 
found in the literature (e.g., \citealt{edvardsson93,casagrande11}). 

The addition of radial or spatial velocities for these objects provides an approach 
to study dynamical processes within the disk, such as radial migration and 
kinematical heating (e.g., \citealt{haywood08}). Local stellar 
samples with both kinematical and chemical information may also be used to 
identify stars from different Galactic components within the thin-disk dominated 
population in the solar neighborhood (\citealt{gratton03,bensby06,karatas12}). 
In order to improve the accuracy of these studies and to extend them over larger 
volumes, massive spectroscopic surveys such as SEGUE (\citealt{yanny09}) and 
RAVE (\citealt{steinmetz06}) have been developed, with additional projects 
expected in the near future (e.g., GALAH; \citealt{zucker12}). 

Although its main scientific goal is the study of the formation and evolution 
of giant planets, brown dwarfs (BDs) and low-mass stars, it was soon realized that 
the Multi-object APO Radial Velocity Exoplanet Large-area Survey (MARVELS; 
\citealt{ge08,ge09};\citealt{geis09}) could also contribute to a better understanding 
of the chemical and kinematical evolution of the solar neighborhood.
During its four-year operation (2008 - 2012) as part of the third phase of the Sloan 
Digital Sky Survey (SDSS-III; \citealt{york00,eisenstein11}), a sample of 
$\sim$3,300 FGK stars with 7.6 $\leq$ V $\leq$ 12.0 (many of which were never previously
analyzed) had their radial velocities (RVs) monitored in the search for companions. 
The targets were selected according to a limited number of well-defined criteria that 
did not explicitly include any cuts based on the metallicities, 
activity levels and ages of the stars (for more details, see \citealt{lee11}). The final 
sample thus presents a well characterized selection function, providing a statistically 
homogeneous data set.

The kinematical analysis will benefit from the precise RVs delivered by 
the MARVELS survey ($\sigma_{RV} \lesssim$ 100 m s$^{-1}$). The chemical analysis, on 
the other hand, depends on the availability of precise atmospheric parameters (effective 
temperature, \teff, and surface gravity, \logg) and chemical abundances (metallicity, 
[Fe/H], and $\alpha$-element content, [$\alpha$/Fe]). 
Stars with RV variations suggestive of the presence of companions were selected for a 
more detailed study, which included the acquisition of high-resolution 
(R $\gtrsim$ 30,000) spectra to determine their fundamental parameters by applying 
standard spectroscopic techniques (excitation and ionization equilibria of Fe I and Fe II 
lines; see the details in \citealt{wisniewski12}).

Stars without detected companions, which correspond to $\sim$80\% of the survey targets, 
were not be subjected to a similar high-resolution spectroscopic follow-up and thus the 
stellar characterization had to rely solely on the MARVELS data. Although
the numbers of visits for each target (typically $\gtrsim$20) were able to 
produce final combined spectra with high signal-to-noise ratios (S/N $\gtrsim$ 100), 
the moderate resolution (R $\sim$ 12,000) and somewhat limited wavelength range 
($\sim$5000 - 5700 \AA) of the data prevented the usage of the classical 
analysis mentioned above. In this resolution regime, most of the lines are blended with 
neighboring features and the stellar characterization through the measurement of 
equivalent widths (EWs) of individual iron lines is not feasible.

Recent large spectroscopic surveys operating in the low-to-intermediate resolution 
($\sim$2,000 -- 20,000) regime have developed pipelines that rely exclusively, or at 
least partially, on the spectral synthesis technique in which the stellar parameters 
are determined through a comparison of the observed spectra with an extensive library 
of previously calculated synthetic ones. As examples, we can mention the pipelines 
from SEGUE (SSPP; \citealt{lee08,smolisnki11}), RAVE (\citealt{siebert11,kordopatis13}), 
LAMOST (\citealt{wu11}) and AMBRE (\citealt{delaverny12,worley12}). 
Although spectral fitting is a powerful technique and provides precise results for 
high-quality data (e.g., \citealt{vf05}), it has some drawbacks: (1) dependency on a 
very detailed and extensive list of spectral lines (many of which may have poorly
determined atomic parameters); (2) the need to accurately know the broadening 
parameters (instrumental profile, macroturbulence and rotational velocities).
(3) stronger correlations (relative to the excitation and ionization equilibria method)
between the resulting atmospheric parameters (see, e.g., 
\citealt{torres12}). For the specific case of MARVELS, the relatively small wavelength 
coverage ($\Delta \lambda \simeq$ 700 \AA) combined with the moderate resolution can 
significantly decrease the accuracy of the spectral synthesis technique. Therefore,
a different method is necessary to efficiently and accurately derive the parameters of 
the stars that did not have high-resolution follow-up spectra.

The purpose of this work is to develop and validate an alternative approach based on spectral 
indices -- specific spectral regions combining multiple absorption lines into broad, blended 
features -- to determine atmospheric parameters directly from the MARVELS spectra, without
any other priors. Indices have been successfully applied before to derive information 
on mean stellar ages and metallicities of populations of galaxies and stars (e.g., 
\citealt{worthey94,trager00,sanchez07,ogando08}), as well as atmospheric parameters 
for target selection purposes in planet search surveys (e.g., 
\citealt{robinson06,robinson07}). Because a manual object-by-object analysis of a numerous 
sample such as that from MARVELS is prohibitively time-consuming, we have developed 
three codes to automate the determination of stellar parameters.

This paper describes the proposed automatic approach to perform the stellar characterization
of MARVELS targets and validates its results using data from the survey and the ELODIE stellar
library (\citealt{ps01}, \citealt{ps07}; see website\footnote{http://www.obs.u-bordeaux1.fr/m2a/soubiran/elodie\_library.html} 
for the most updated 3.1 version). The paper is organized 
as follows. Section \ref{data} presents the data used to build the calibrations that were 
later adopted for the derivation of the atmospheric parameters. A sample of MARVELS
stars utilized to test the precision of the results is also described. 
The definition of the spectral indices is detailed in Section \ref{indices}, while Section 
\ref{pipeline} is devoted to a thorough description of the method. The four 
steps of our analysis (normalization of the spectra, measurements of EWs, construction of 
the calibrations and derivation of atmospheric parameters) are explained in separate 
subsections. The results obtained with the spectral indices are shown in Section \ref{results}, 
along with a discussion regarding their accuracy. Finally, our concluding remarks are presented 
in Section \ref{conclusions}.

\section{Data}

\label{data}

Two different samples of stars were used in this work. The first is formed by stars with
well-known atmospheric parameters (derived from the analysis of high-resolution spectra) and
was utilized to construct the calibrations used to obtain the atmospheric parameters 
(\teff, [Fe/H] and \logg), thus being referred to as the \textit{calibration sample}. The 
second contains a subset of MARVELS targets for which independent high-resolution spectra and 
precise atmospheric parameters are available. As its purpose was to check the performance of 
the spectral indices approach, we called it \textit{validation sample}.
Both samples are described in more detail below.

\subsection{Calibration Sample}

\label{calibsample}

\begin{figure}
\plotone{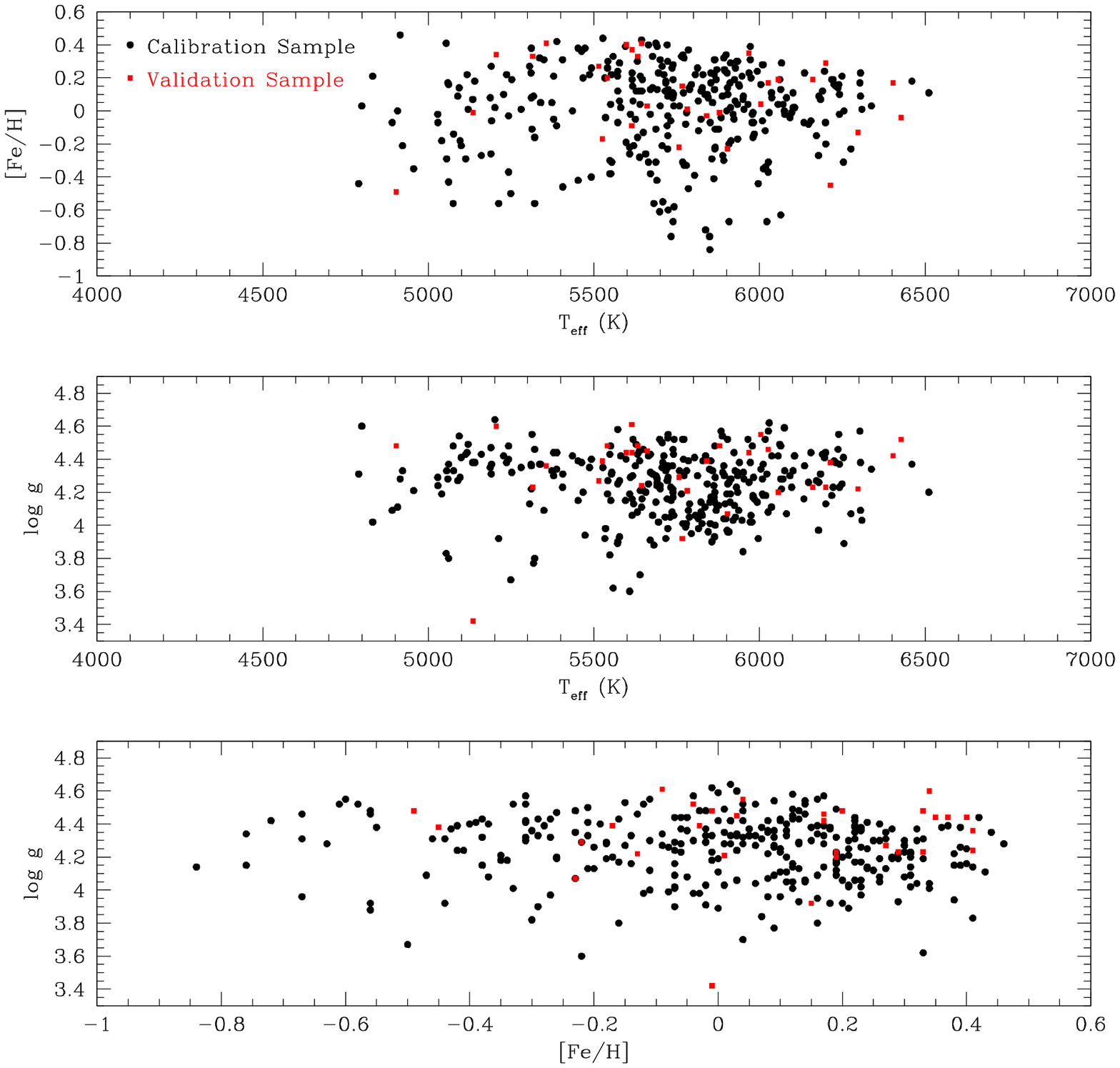}
\caption{The atmospheric parameters (\teff, [Fe/H] and \logg) for the 309 stars in
the calibration sample (black circles) and the 30 stars in the validation sample 
(red squares). The intervals covered by the calibration sample are: 4800 K 
$\lesssim$ \teff\, $\lesssim$ 6500 K, 3.60 $\lesssim$ \logg\, $\lesssim$ 4.70 and 
$-$0.90 $\lesssim$ [Fe/H] $\lesssim$ $+$0.50. The calibration stars also cover
the parameter space of the MARVELS sample, represented here by the validation sample.}
\label{samples}
\end{figure}

The calibration sample is composed of 309 stars; their high-resolution spectra,
effective temperatures, surface gravities and metallicities were taken from 
\cite{ghezzi10a,ghezzi10b} and \cite{peloso05a,peloso05b}. Figure \ref{samples} 
shows the distribution of these stars (open circles) in parameter 
space; it is clear that this sample has a good coverage in the 
following intervals: 4800 K $\lesssim$ \teff\, $\lesssim$ 6500 K, 3.60 $\lesssim$ 
\logg\, $\lesssim$ 4.70 and $-$0.90 $\lesssim$ [Fe/H] $\lesssim$ $+$0.50.


The majority of the stars were taken from \cite{ghezzi10a,ghezzi10b}, but the 17 stars 
classified as giants by the authors were not included here. This cut was done because 
their spectra are considerably different from those of dwarfs and subgiants and a proper
analysis would require a distinct set of spectral indices. 
The remaining 291 stars (262 dwarfs and 29 subgiants) all have high-resolution (R $\sim$ 
48,000) and high-quality (S/N $\gtrsim$ 200 per resolution element at $\lambda \sim$ 6700 
\AA) spectra obtained with the Fiber-fed Extended 
Range Optical Spectrograph (FEROS) spectrograph (\citealt{kaufer99}) attached to the 
MPG/ESO 2.2 m telescope (La Silla, Chile), under the agreement between ESO and 
Observat\'orio Nacional (MCT). Additional details about the sample selection, 
observations and data reduction are given in \cite{ghezzi10a}. The authors also describe 
the homogeneous derivation of the atmospheric parameters (\teff\, and \logg) and 
metallicities ([Fe/H]) that were utilized in this work. In summary, the standard
spectroscopic analysis based on the excitation and ionization equilibria of a
carefully selected list of Fe I and Fe II lines was used to determine the results
in an automated procedure.

In order to better populate the region [Fe/H] $< -0.50$ of the parameter space, the above 
sample was complemented with stars from \cite{peloso05a,peloso05b}. Nine stars 
are common with \cite{ghezzi10a,ghezzi10b} and were considered only for comparative purposes 
(see below). The remaining 18 stars have high-resolution (R $\sim$ 48,000) and high-quality 
(S/N $\gtrsim$ 300) spectra acquired with the FEROS spectrograph fed by the ESO 1.52 m telescope. 
The details about the sample selection, observations and data reduction for these objects can be 
seen in \cite{peloso05a}. A description of the iterative method used to determine the 
atmospheric parameters (\teff\, and \logg) and metallicities ([Fe/H]) adopted here are also 
provided by the authors. Briefly, the effective temperatures were estimated from the arithmetic 
mean of the values derived from photometric calibrations and H$\alpha$ profile fitting. 
The surface gravities were derived using these effective temperatures and also stellar 
luminosities and masses. The luminosities were calculated from Hipparcos parallaxes and V magnitudes.
The masses were obtained from interpolation in grids of evolutionary tracks using the effective
temperatures, luminosities and metallicities. Finally, the metallicities were determined from a 
differential analysis relative to the Sun using Fe I and Fe II lines. 

The 291 stars from \cite{ghezzi10a,ghezzi10b} and 18 from \cite{peloso05a,peloso05b} compose the
final calibration sample used in this work. Although the two subsamples were analyzed with somewhat 
different methods, we find that the resulting parameters are consistent. 
For the 9 stars in common, the average differences are (in the sense Ghezzi - del Peloso): 
$\Delta$\teff\, = $-$4 $\pm$ 41 K, $\Delta$[Fe/H] = $-$0.04 $\pm$ 0.04 dex, 
and $\Delta$\logg\,= 0.00 $\pm$ 0.12 dex. In this paper, the spectra and parameters from 
\cite{ghezzi10a,ghezzi10b} were adopted for these nine stars.
We note that these parameters are consistent with those from many other catalogs of atmospheric 
parameters available in the literature (see, e.g., Table 4 of \citealt{ghezzi10a}).

To ensure consistency between the calibration and MARVELS survey data (the latter is 
described in Section \ref{testsample}), the FEROS spectra were degraded and resampled to 
the MARVELS resolution and sampling. The IRAF\footnote{IRAF (Image Reduction and Analysis 
Facility) is distributed by the National Optical Astronomy Observatories (NOAO), which is 
operated by the Association of Universities for Research in Astronomy, Inc. (AURA) under 
cooperative agreement with the National Science Foundation (NSO).} tasks \texttt{gauss} 
(with a value of 8.0 for the \textit{sigma} parameter) and \texttt{dispcor} (with a value 
of 0.154 for the \textit{dw} parameter) of  were used, respectively, to accomplish this. 
The spectra were also trimmed with IRAF's task \texttt{splot} in order to keep only the 
region 5100 - 5590 \AA, which is present in the majority of the MARVELS spectra.  

\subsection{Validation Sample}

\label{testsample}

The validation sample consists of the full set of 30 stars that currently have both low resolution 
MARVELS spectra and precise atmospheric parameters resulting from the analysis of high-resolution 
spectra. Sixteen of these stars are MARVELS targets with detected companions which were or are 
currently being subjected to more detailed analyzes\footnote{The MARVELS 
candidate MC10 (TYC 3010-1494-1; \citealt{mack13}) was not included in the validation sample, 
despite having a set of high-resolution parameters, because it is the primary component 
of a binary system with a mass ratio close to 1.}. Their high-resolution spectra were 
obtained with the ARC Echelle Spectrograph (ARCES; \citealt{wang03}) attached to the 
Astrophysical Research Consortium (ARC) 3.5 m telescope at the Apache Point Observatory 
(APO; New Mexico, USA) and/or the FEROS spectrograph attached to the MPG/ESO 2.2 m telescope. 
The resolutions are $\sim$31,500 and $\sim$48,000, respectively, and all spectra have 
S/N $\gtrsim$ 100 per resolution element. 
Their atmospheric parameters were derived following the method described in \cite{wisniewski12} 
and are given in Table \ref{marvels_cand_param}. Briefly, they were derived from two 
independent pipelines, both based on the technique of excitation and ionization equilibria of 
Fe I and Fe II lines. Given the consistency of the results, the two sets of parameters were combined 
through a weighted average, using the internal uncertainties from each pipeline as the weights.
 
The other 14 stars are known planet-hosts 
which were used in the MARVELS survey as reference objects for the RV determinations. 
Their atmospheric parameters are the arithmetic average of the values taken from multiple 
sources in the literature (see Table \ref{marvels_ref_param}).  
In this table, the uncertainties correspond to the standard deviations of the average values
from the literature (and not to internal errors of the method, as in Table 
\ref{marvels_cand_param}). The distribution of parameters for the thirty stars in the 
validation sample is also presented in Figure \ref{samples} (filled squares).

The low resolution spectra for the validation sample were obtained, as part of the survey, 
with the MARVELS instrument (\citealt{ge08,ge09}; \citealt{geis09}) coupled to the SDSS 2.5 m 
telescope at APO (\citealt{gunn06}). The MARVELS instrument is a 60 object, fiber-fed, Dispersed 
Fixed Delay Interferometer (DFDI; \citealt{wang12a,wang12b}) that outputs two fringing spectra 
(``beams'') per object with a resolution R $\sim$ 12,000 and a wavelength coverage between 
$\lambda\lambda \sim$ 5000 - 5700 \AA. Since each star was visited typically $\gtrsim$ 20 times, 
the final combined spectra have S/N $\gtrsim$ 100 per pixel (dispersion 0.154~\AA/pixel).
To convert the fringing spectra to conventional 1D extracted spectra, the former are first 
preprocessed, including corrections for optical distortion and slit illumination, creating 
a continuum-normalized fringing spectrum. Then, for each wavelength, a sinusoid is 
fit to the fringing pattern which lies perpendicular to the wavelength axis. The DC offset 
of this sinusoid is the flux value of the normalized 1D extracted spectrum for that wavelength. 
These fluxes (counts per pixel) are converted into flux densities (counts per unit wavelength), 
and then a barycentric RV correction is applied such that all 1D extracted spectra for a given 
star are registered to the rest wavelength.

The determination of atmospheric parameters from these low-resolution MARVELS spectra and 
spectral indices is described in the following sections. The ability of the method to 
accurately recover the high-resolution results previously derived for the MARVELS stars is 
regarded as the validation test of the approach presented in this paper. We note that, in spite of 
its relatively small size, the adopted validation sample offers the most realistic check for the 
quality and reliability of our method since it uses real MARVELS data (which includes instrumental 
effects, noise, etc). Moreover, the set covers most regions of the parameter space defined by the 
calibration sample (see Figure \ref{samples}) and also the ranges of atmospheric parameters expected 
for the whole MARVELS sample (according to the selection criteria employed during the survey 
target selection).

\section{Definition of the Spectral Indices}

\label{indices}

As already mentioned, the MARVELS resolution prevents the characterization of the stars 
through measurements of equivalent widths of individual lines (and consequently classical
model atmosphere analysis) because most of the spectral lines become blended with 
neighboring features. To overcome this problem, 
we have used instead an approach based on spectral indices 
(e.g.,\citealt{worthey94,robinson06}), which are defined here as 
groups of lines formed by similar chemical species. More specifically, we have 
selected two groups of interest with features dominated by: (1) neutral iron-peak 
species (such as Fe I, V I, Cr I, Mn I, Co I and Ni I) and (2) ionized species (such 
as Fe II, Ti II, and Cr II). These groups have, in principle, properties that 
should allow us to constrain \teff, [Fe/H] and \logg. Note that we consider 
a feature to be dominated by a certain group of elements when their respective 
lines account for more than 90\% of its total absorption.

\begin{figure}
\epsscale{1.0}
\plotone{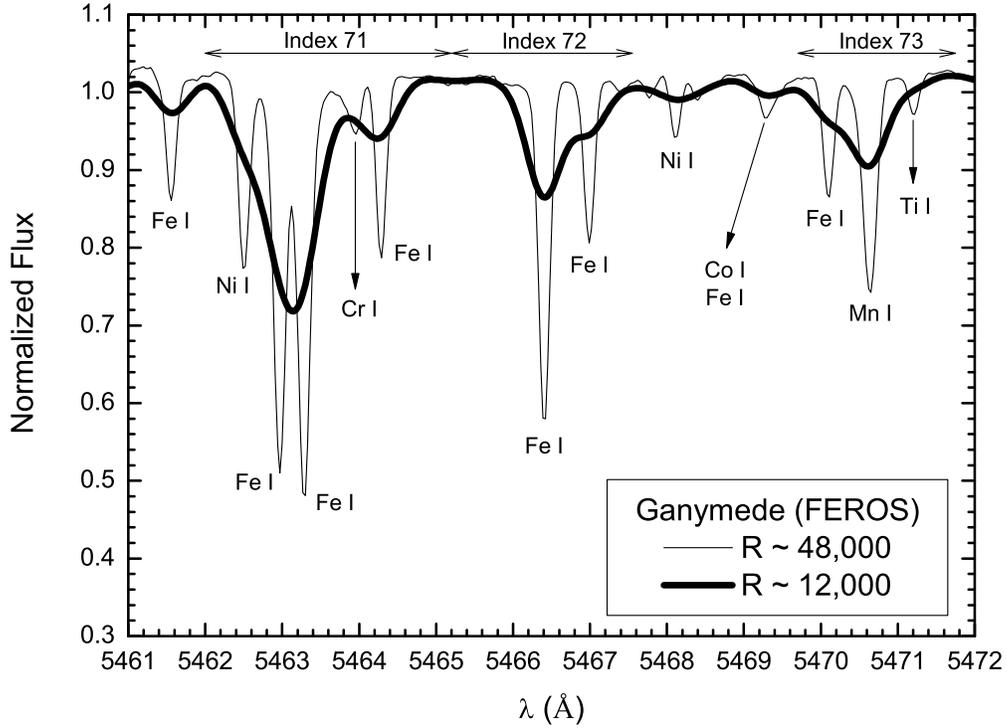}
\caption{Section of the FEROS Ganymede spectrum used as a solar template. The 
thin line shows the spectrum at its original resolution (R $\sim$ 48,000)
and the thick line represents the version that was degraded to the MARVELS
resolution (R $\sim$ 12,000). The most prominent individual lines are identified
by their chemical species. Three spectral indices dominated by neutral iron-peak 
species and their respective wavelength intervals of are shown as examples of the 
selection procedure.The other indices are distributed throughout the full wavelength 
range used in the analysis (5100--5590~\AA).}
\label{indices_plot}
\end{figure}

The two groups were identified through a detailed inspection of the interval 
5100 - 5590~\AA\, on two versions of the same FEROS Ganymede spectrum 
(\citealt{ribas10}) that was adopted as a solar template 
(see Figure \ref{indices_plot}). The first version 
retained the original resolution (R $\sim$ 48,000) and was used in conjunction with 
\cite{moore66} to identify the individual lines that composed the indices. 
The second version was degraded to the MARVELS resolution and sampling (see section 
\ref{calibsample}) and was used to determine the initial and final wavelengths of 
the selected indices.


\begin{figure}
\epsscale{1.0}
\plotone{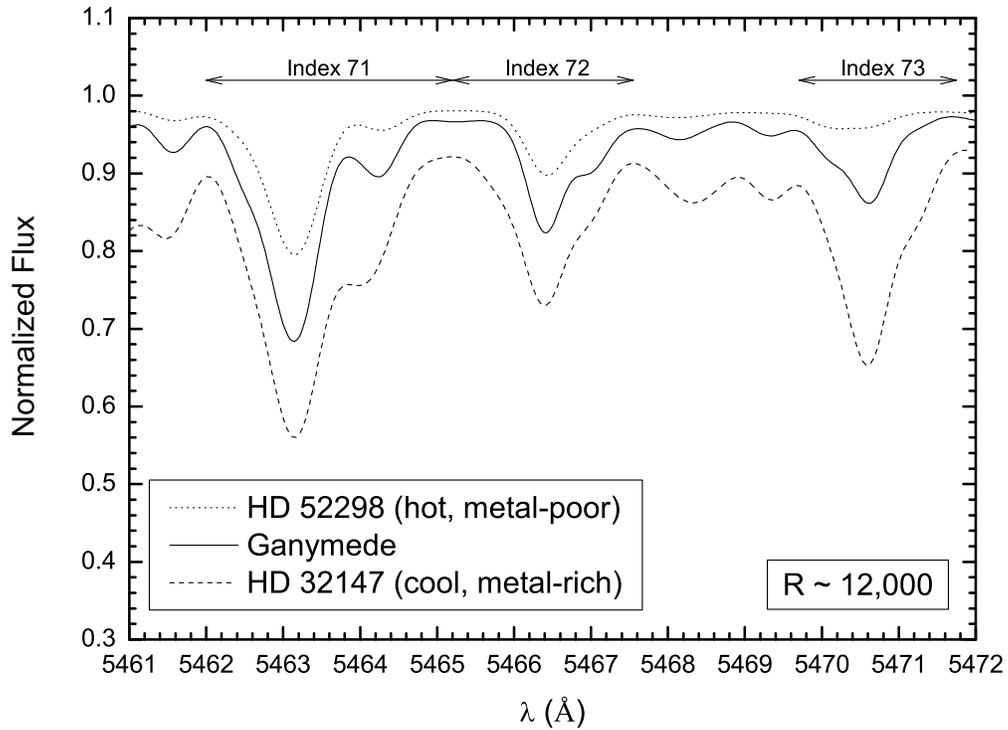}
\caption{Section of the FEROS spectra of HD 52298 (dotted line), Ganymede (solid
line) and HD 32147 (dashed line). They were all degraded to the MARVELS
resolution (R $\sim$ 12,000) and the continua have been shifted vertically for 
clarity. The three spectral indices shown as examples are the 
same as in Figure \ref{indices_plot}. It is clear that their definitions are 
appropriate for the three stars, although their parameters are significantly
different.}
\label{indices_three_stars}
\end{figure}

Degraded FEROS spectra of the stars HD 32147 and HD 52298 were also utilized during 
the selection of the indices (see Figure \ref{indices_three_stars}). HD 32147 is a 
cool metal-rich star (\teff\, = 4850 K and [Fe/H] = 0.25; \citealt{silva07})
and corresponds to an extreme example of a spectrum with very strong indices.
The analysis of its spectrum allowed us to refine the wavelength intervals previously 
defined for the indices (based on the solar spectrum) and to eliminate those that 
were not sufficiently isolated (i.e., did not have clear apparent continuum
regions between them and the neighboring features).
HD 52298 is a hot metal-poor star (\teff\, = 6253 K and [Fe/H] = $-$0.31; 
\citealt{peloso05a}) and illustrates the opposite case, in which the strengths of 
the indices become too weak, preventing accurate measurements of their equivalent 
widths. In summary, these two stars probe the extremes of the anticipated absorption 
strengths of the spectral indices and helped us to exclude possibly problematic features.


Following the procedure described above, a total of 96 potential indices 
were selected: 80 dominated by neutral iron-peak species and 16 by ionized 
species. The properties of these indices are listed in Table \ref{indices_properties}.
We have referred to them as potential indices because their actual 
sensitivities to the atmospheric parameters are only analyzed in Section
\ref{regressions}. The final list with the best available indices for the subsequent 
characterization of the MARVELS sample resulted from this analysis.

\section{The Spectral Indices Method}

\label{pipeline}

The MARVELS sample contains $\sim$3,300 stars with a high enough number of visits for
robust planet detection and the time required to manually
analyze all objects, each with $\gtrsim$40 spectra, would be prohibitively large. 
Moreover, such an approach would be more error prone due to subjective choices
that are inevitably made throughout the analysis of stellar spectra (e.g., during the 
continuum normalization). 
To avoid these issues, the determination of the atmospheric parameters was automatized 
through the development of three codes, which will be made publicly available
throught the Brazilian Participation Group Scientific Portal
\footnote{http://bpg.linea.gov.br/}. 
These codes, as well as the tests performed to ascertain
their quality, are described in more detail in the following sections.

\subsection{Normalization}

\label{norm}

The normalization of a spectrum is a time-consuming 
task since one must test several parameters for choosing the curve that best 
describes the continuum of a particular star. As we aim to apply the indices 
technique for large stellar samples (in particular, from the MARVELS survey), an automatic 
tool to perform this task is required. To this end, the first program 
was developed to fit a curve to and normalize the continuum on the spectra of solar-type 
stars. It was inspired by the task \texttt{continuum} from IRAF, but has the 
advantage of automatically testing many different normalizations and 
determining which one provides the best fit. This feature turns the normalization 
process into a fast, non-interactive and systematic procedure, characteristics that 
are essential for the analysis of large samples of stars.

The code uses as input the reduced, defringed, 1D Doppler-corrected spectra 
(in FITS format) that are given as part of the final products delivered by MARVELS  
pipeline. It is possible to combine more than one spectrum using the median counts 
as weights and cosmic rays can be removed through a $\sigma$ clipping algorithm 
if at least five spectra are being co-added. A number of 1D Legendre polynomials are 
then fit to the continuum points of the individual or combined spectra, in a wavelength 
range defined by the user. Both the number of polynomials tested and of points 
considered in the fit depend on an additional set of input parameters provided by 
the user: polynomial order, high and low rejection limits (as well as 
the steps with which they are changed), grow parameter and number of iterations
to be performed. All parameters have the same definitions as in the task 
\texttt{continuum} from IRAF.

The program computes as many solutions as the choice of the input parameters described 
above, but testing all possible combinations for each star would require an excessive amount 
of time and most of the solutions would not be considered satisfactory. Thus, we decided to restrict 
the set of input parameters using 27 stars (seven with degraded FEROS spectra and 20 MARVELS 
targets with spectra from the survey) with parameters that sample the range of indices strengths 
expected to be found in the stars from the MARVELS survey. We note that this test set
included cool metal-rich stars to ensure that suitable apparent continuum regions could still be 
found, allowing accurate normalizations. The best visual normalizations
were found for the following values of the input parameters: 5 or 6 for the
order of the Legendre polynomials; 1.0 to 1.5 and 3.0 to 4.0 (both with steps of 0.5)
for the low and high rejection parameters; 1 for the grow parameter; 11 for the number 
of iterations. It was also observed that the fits with the highest correlation 
coefficient $R^{2}$ always produced the best results (the same was not true for the 
cases with the lowest values of the standard deviation of the residuals $\sigma$)
and this is the reason why we chose this statistical criterion to select the final solutions. 

There are some limitations that could 
not be circumvented by the code after these restrictions were made. These issues are
mainly related to the inability to deal with cosmic rays (due to inefficient 
clipping or insufficient number of spectra to be combined) or regions with 
defects (e.g., strong curvatures on the edges). The former can be minimized 
by combining a larger number of spectra for each star (which was not possible
for the stars with FEROS data). The second limitation is more complicated, but 
the usage of higher values for the polynomial order proved to be effective
in most of the cases. 

\begin{figure}
\epsscale{0.7}
\plotone{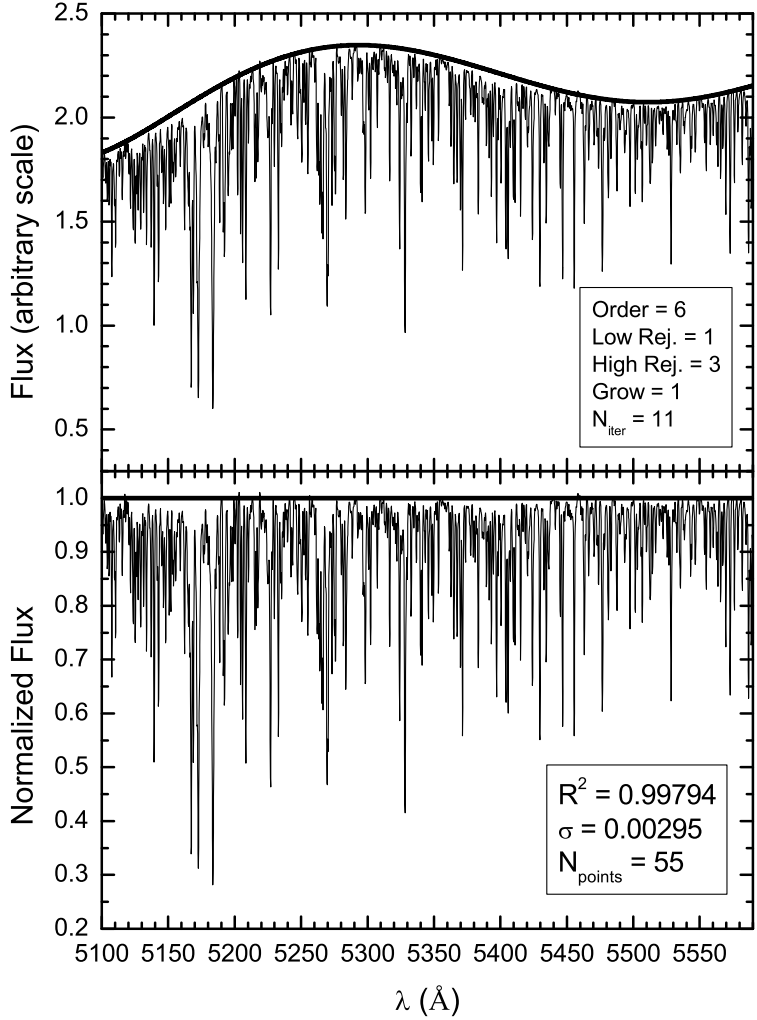}
\caption{Example of the normalization procedure. \textit{Upper panel:} Degraded FEROS 
spectrum of the calibration star HD 20630 before the normalization. 
The thick black line represents the polynomial that best fits the continuum. 
The values of the polynomial order, low and high rejection limits, grow parameter and 
number of iterations are provided. \textit{Lower panel:} Same spectrum after the
continuum normalization. The black thick line now shows the normalized 
continuum level at 1.0. The correlation coefficient $R^{2}$, standard deviation $\sigma$ 
and number of points of the fit are also given.}
\label{examples_norm}
\end{figure}

The spectra of the 309 stars from the calibration sample (Section 
\ref{calibsample}) were normalized using the code and the restrictions and criteria
described above. An example of the normalization procedure is shown in Figure 
\ref{examples_norm}. We visually inspected the final solutions for each of the 309 
stars to search for normalization problems that could undermine the indices 
calibrations. The solutions provided by the code had to be replaced by visually better 
fits in 9\% of the cases, which defines a reasonable limit to the degree of efficient 
automation of the normalization process. For the analysis of the complete MARVELS sample,
the number of spectra will be larger by a factor of $\sim$100 and such a visual inspection
will not feasible. We are currently working on a complementary code to automatically identify 
these cases that require human intervention based on large variations of the continuum fit 
parameters or the measured EWs for some indices.

Even though an effort was made to ensure that the continuum placement was as accurate as 
possible, it is possible that small offsets are still present for a few spectra. The 
typical S/N of the individual MARVELS spectra for the validation sample is $\sim$100, 
which corresponds to a rms error of the continuum of $\sim$(S/N)$^{-1}\sim$0.01.
Using the method described in Sections \ref{ews}, \ref{regressions} and \ref{method} as 
well as the final results presented in Section \ref{finalresults}, we have checked that an 
offset of $\pm$1\% in the continuum placement causes the following average variations on the 
derived atmospheric parameters: $\Delta$\teff = $-$5 $\pm$ 42 K, $\Delta$[Fe/H] = $\pm$0.12 $\pm$ 0.04 
and $\Delta$\logg = $\mp$0.07 $\pm$ 0.07. Note that these values correspond to the unlikely case
in which a global systematic offset would shift all the spectra of all the stars in the sample 
in the same direction and by the same amount. Therefore, they should be regarded as the maximum possible
errors that could be introduced by normalization issues. The actual errors will therefore be smaller 
than estimated here (see also discussion in Section \ref{finalresults}).


\subsection{Measurements of the Equivalent Widths of the Indices}

\label{ews}

The normalized spectra obtained above for the calibration stars were used as input 
for the next code, which measures the equivalent widths (EWs) of a list of indices 
provided by the user by direct integration of their profiles. This method was chosen
because most of the indices are formed by multiple lines, each with a different type
of profile (Gaussian or Voigt). The integrations were performed in the wavelength 
intervals defined for each index (see Table \ref{indices_properties}) adopting
the value of 1.0 for the normalized continuum flux. 
 
This code was used to measure the equivalent widths of the 96 indices (see Section 
\ref{indices}) in the degraded FEROS spectra of all 309 stars from the calibration
sample. We have checked that all indices' EWs could be correctly measured even for the 
most metal-poor stars in the sample. A quick look at columns ``EWmin'' in Table 
\ref{calibrations_ew} and ``Notes'' in Table \ref{indices_properties} reveals that 4.81 m\AA\,
is the lowest value for a equivalent width used in the calibrations that were adopted for the 
derivation of the atmospheric parameters (more details about this selection are given in the 
next section). This value occurs for the star HD 76932 (\teff = 5850 K and [Fe/H] = -0.84) 
in the calibration for index 80 and is consistent with the lower limit of the measuring capabilities 
of our automated software.

Problematic EWs were identified and discarded during the construction of the regressions 
(see Section \ref{regressions}), since bad measurements appeared as clear outliers 
relative to the general trends observed. For the MARVELS targets, multiple spectra from
different visits are available for a given star and the outliers can be excluded 
by a simple $\sigma$ clipping procedure.

\subsection{Construction of the Calibrations}

\label{regressions}

The previous section discussed the measurements of the equivalent widths of the 96 
spectral indices in the spectra of the calibration stars.
It was also mentioned in Section \ref{calibsample} that all these objects 
have precise atmospheric parameters (\teff, [Fe/H] and \logg) derived from detailed 
and homogeneous analysis of high-resolution spectra. 
With both the EWs and atmospheric parameters in hand, we explored the relations 
between these quantities through a multivariate analysis and the final result
was a set of calibrations that allowed the subsequent characterization of the  
MARVELS validation sample based solely on spectral indices (without any other priors).

\begin{figure}
\epsscale{1.0}
\plotone{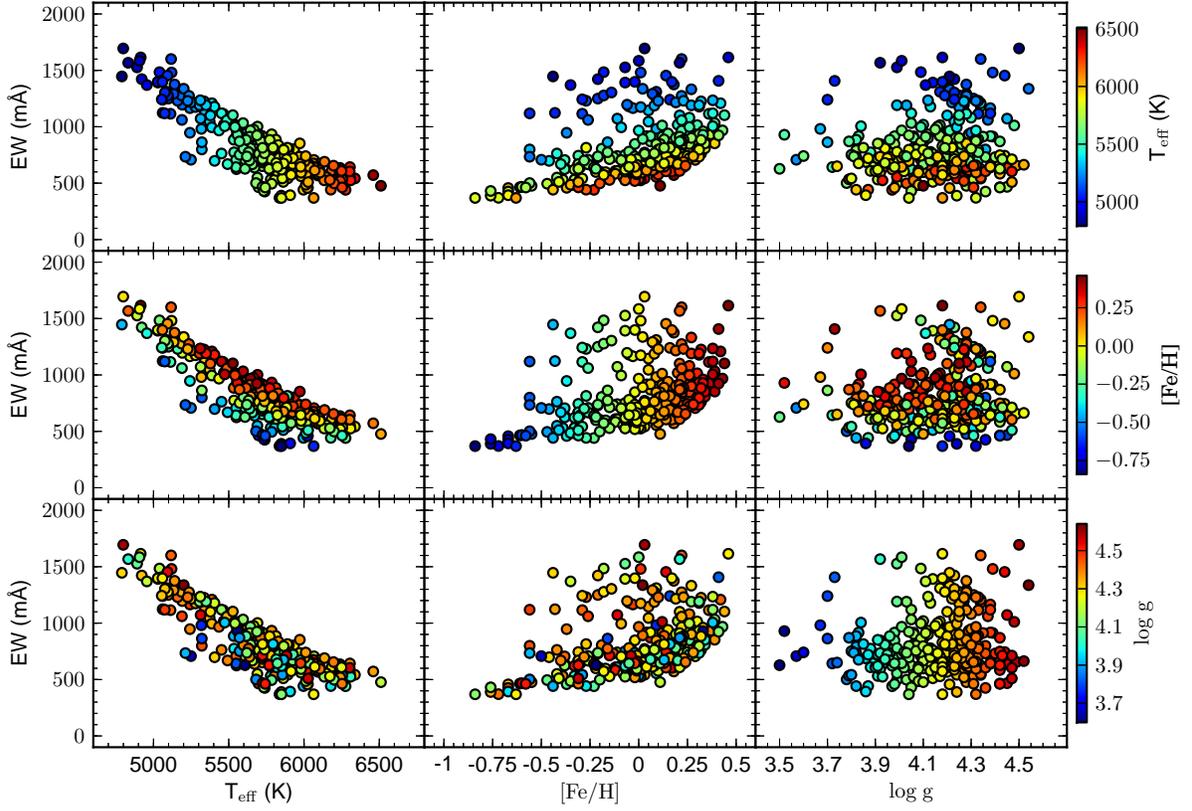}
\caption{Measured equivalent widths of index 38 (see Table \ref{indices_properties}) as a
function of the atmospheric parameters. The left, middle and right columns show the
dependencies with \teff, [Fe/H] and \logg, respectively. In the upper, middle and lower
rows, the points are color-coded according to their effective temperatures, 
metallicties and surface gravities, respectively. The EWs exhibit significant correlations 
with \teff\, and [Fe/H], but no visible dependency on \logg\, is observed.}
\label{index_sensitivity}
\end{figure}

Figure \ref{index_sensitivity} shows an example of the behavior of the indices' EWs as
a function the atmospheric parameters for the calibration sample. Clear relations are readily 
visible for \teff\, and [Fe/H], while only scatter can be seen for \logg. Although this
initial qualitative analysis revealed some promising indices, it can be misleading for the 
cases which lack evident correlations. The scatter might result from a truly weak sensitivity of 
the index relative to the parameter in question, but can also be caused a stronger sensitivity to 
one of the other two parameters (recall that these plots show only a 1D projection of a possibly 
3D relation). Problems during the measurement of the EWs for particular indices (e.g., possible 
contaminations having larger contributions than anticipated) certainly also play a role. Therefore, 
we decided not to remove any of the indices based on this initial visual inspection. 


The starting point of the quantitative analysis was the choice of an appropriate model
to describe the relations observed in plots similar to Figure \ref{index_sensitivity}.
In principle, all indices' EWs depend on the three atmospheric parameters in a complicated 
way which includes the interdependency between these parameters. Tests with different
polynomial orders revealed that a quadratic model would be able to correctly
describe most of the observed behaviors. Thus, we decided to search for the best 
calibrations for each index using second-order polynomials with the following structure: 

\begin{eqnarray}
\label{ewcalib}
{\rm EW\, (m\mbox{\AA})} = c_{0} + c_{1}{\rm [Fe/H]} + c_{2}T_{\rm eff} + c_{3}\log g + \nonumber \\
c_{4}{\rm [Fe/H]}T_{\rm eff} + c_{5}{\rm [Fe/H]}\log g + c_{6}T_{\rm eff}\log g + \nonumber \\
c_{7}({\rm [Fe/H]})^{2} + c_{8}(T_{\rm eff})^{2} + c_{9}(\log g)^{2}
\end{eqnarray}

These models ensure that any eventual interdependence between the atmospheric parameters
are taken into account. Some of the terms, however, could be statistically 
insignificant and, to evaluate this possibility, the choice of the final best calibrations 
followed an iterative procedure, which was performed with a Python routine developed for 
this work. The starting point was Equation (\ref{ewcalib}), 
for which the coefficients $c_{0}$, ..., $c_{9}$ were determined through the Ordinary 
Least Squares (OLS) method (i.e., without any weights). We have designated these the complete 
models. When the best fit was found, the outliers were removed with a 
2$\sigma$ clipping. The previous steps were repeated three times and this limit was 
chosen to avoid an excessive exclusion of points. The mean number of stars that appeared as 
outliers was 37, with minimum and maximum values of 19 and 49, respectively. These numbers 
correspond to $\sim$6-16\% of the complete calibration sample, which is a reasonable fraction.
We have checked that these outliers do not correspond to the stars that had their 
normalizations replaced by visually better fits (see Section \ref{norm}).
At the end of this iteration, the following statistical 
quantities were obtained for the current model: the correlation coefficient $R^{2}$ of the 
fit, the standard deviation of the residuals ($\sigma$) and the Bayesian Information 
Criterion (BIC). 

The code then searched for models that had the same polynomial order and a similar
ability to fit the data, but with fewer terms; these are the
reduced models. The progression of these tests was hierarchical, i.e., the higher-order 
terms were the first ones to be removed. In each trial, a temporary reduced model and its
associated statistical quantities ($R^{2}$, $\sigma$ and BIC) were calculated. This 
reduced model was then compared to the complete one and the temporarily removed term was 
definitely excluded from the polynomial if three criteria were simultaneously met: 
both $\sigma$ and BIC are lower for the reduced model and the corresponding
coefficient $c_{i}$ is statistically insignificant according to the
F-statistics. If these conditions were satisfied, the reduced model replaced the complete 
one and new trials were done until all terms were tested. If one of the 
conditions was not satisfied, the tested term returned to the model and the trials continued 
until the relevance of all terms was evaluated. After the iterative procedure described
above, a final optimized calibration was obtained for each index.

Table \ref{calibrations_ew} presents the information for the calibrations obtained for the 
total set of 96 indices. If no number is provided for a given coefficient, the corresponding 
term was excluded according to the tests of statistical significance explained above. 
Besides $R^{2}$ and $\sigma$, the number of stars used in the fit and the validity ranges
for the equivalent widths are also shown. The validity ranges for \teff, [Fe/H] and \logg\,
are not given because they are very similar to the parameter space covered by the calibration 
sample (see Section \ref{calibsample}). 

\begin{figure}
\epsscale{1.0}
\plotone{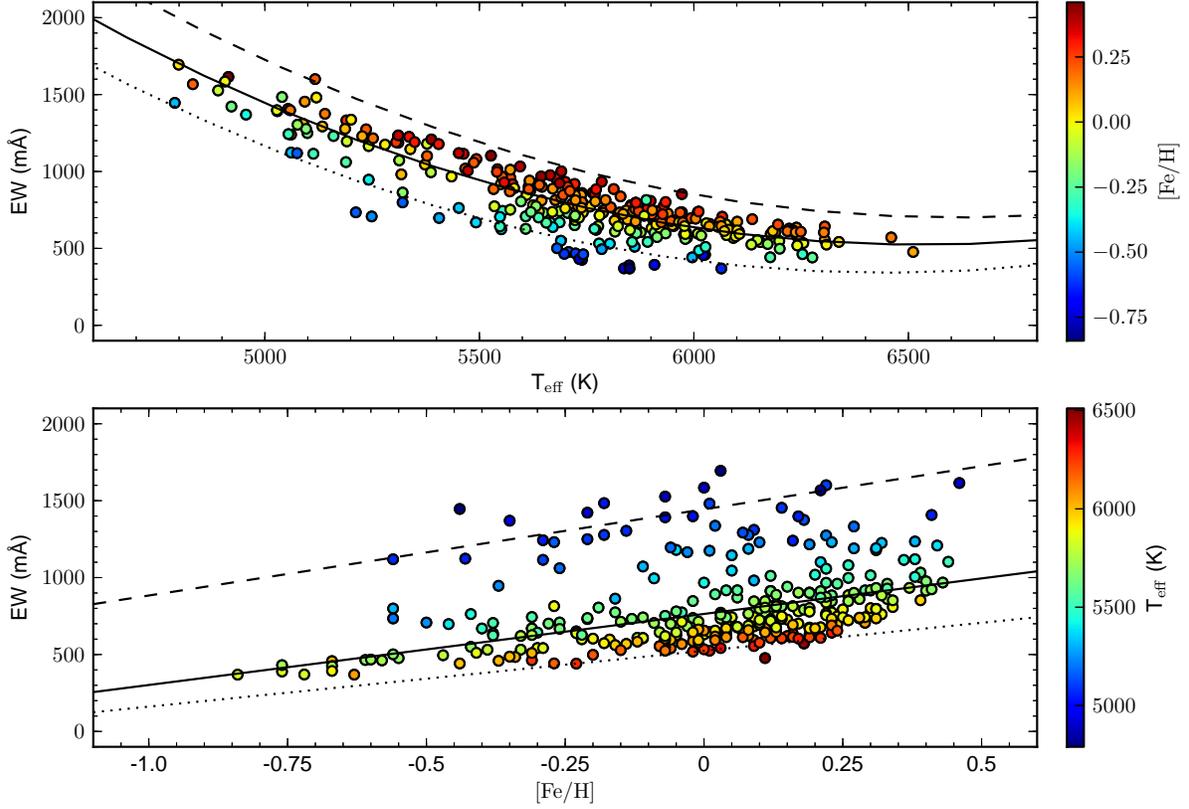}
\caption{Examples of the calibration for index 38 (see Table \ref{calibrations_ew}).
\textit{Upper panel:} Variation of the measured equivalent widths as a function of \teff,
with the points color-coded by [Fe/H]. The dotted, solid and dashed black lines represent
Equation (\ref{ewcalib}) for \logg\, = 4.44 and [Fe/H] = $-$0.50, 0.00 and +0.50, respectively.
\textit{Lower panel:} Variation of the measured equivalent widths as a function of [Fe/H],
with the points color-coded by \teff. The dotted, solid and dashed black lines represent
Equation (\ref{ewcalib}) for \logg\, = 4.44 and \teff\, = 5000, 5750 and 6500 K, respectively.
The calibration accurately represents the observed behavior of the points throughout the
parameter space covered by the calibration sample.}
\label{calib_example}
\end{figure}

In order to guarantee the determination of precise atmospheric parameters, we have decided
to keep in the analysis only those indices for which the respective calibrations had 
$R^{2} \geq$ 0.9. This cut removed four indices (marked with a number 2 in the notes of 
Table \ref{indices_properties}): 1, 41, 64 and 65. Figure \ref{calib_example} shows index 38 
as an example of the final calibrations adopted in this work. 


\begin{figure}
\epsscale{1.0}
\plotone{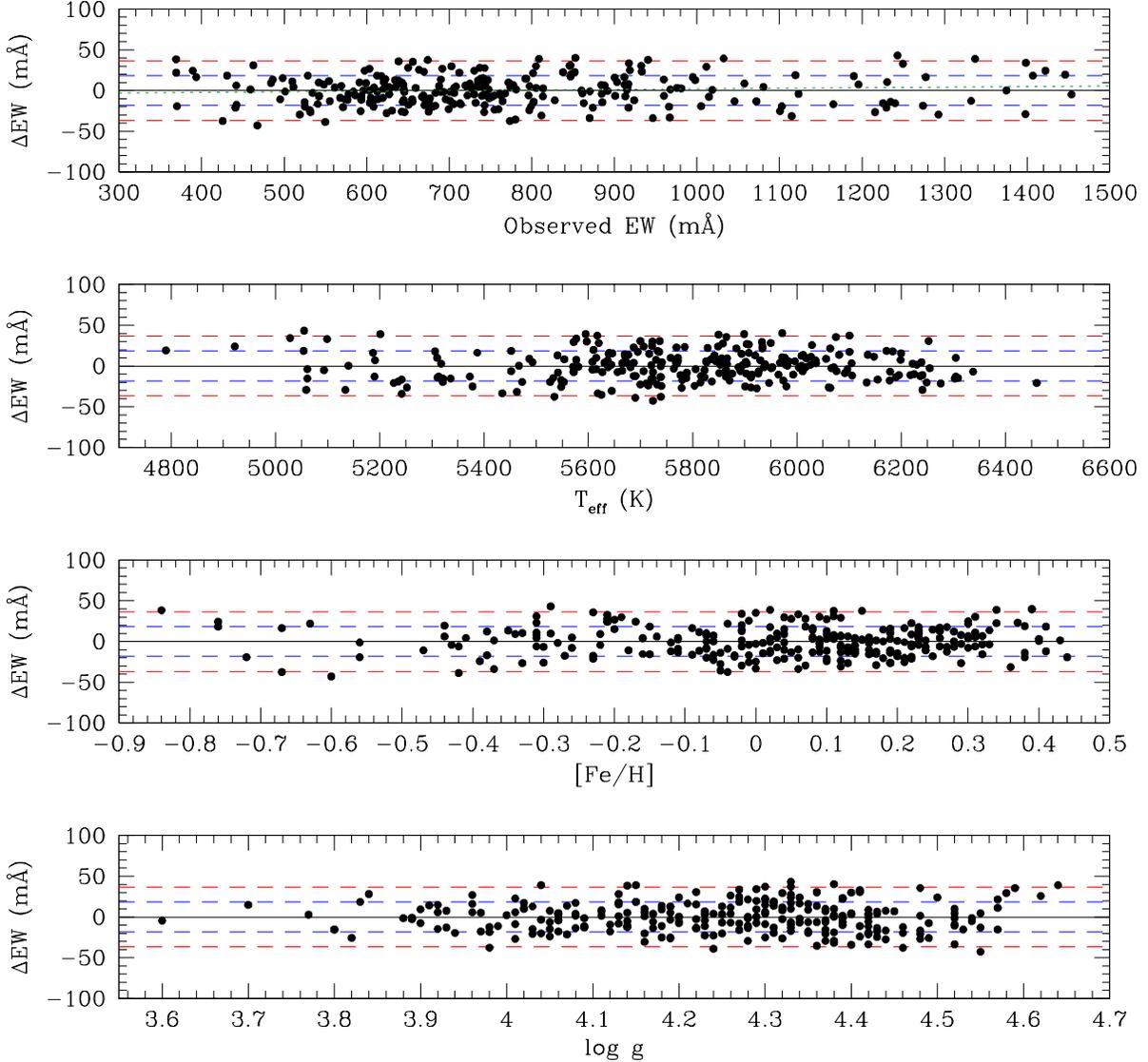}
\caption{Residuals ($\Delta$EW = EW$_{calc}$ - EW$_{obs}$) from the calibrations of index 
38 as a function of, from top to bottom, EW$_{obs}$, 
\teff, [Fe/H] and \logg. The horizontal black solid lines indicate the perfect agreement. The 
blue and red dashed lines show, respectively, the 1 and 2$\sigma$ limits, where $\sigma$ = 18.36
m\AA\, is the standard deviation of the residuals (see Section \ref{calibrations_ew}). The green dotted line 
in the upper panel represents the linear fit to the residuals, which has a correlation coefficient 
$R^{2}$ = 0.007. In the panels for \teff, [Fe/H] and \logg, the fit can not be distinguished from 
the black line. No significant trends can observed for the residuals as a function of the EWs and 
atmospheric parameters.}
\label{residuals}
\end{figure}

The next step was the analysis of the residuals (i.e., the differences $\Delta$EW between 
the equivalent widths EW$_{calc}$ calculated with Equation \ref{ewcalib} and the equivalent 
widths EW$_{obs}$ measured in the stellar spectra) for the remaining 92 calibrations. These were
plotted as a function of the variables (EW, \teff, [Fe/H] and \logg) and linear fits were 
applied to the data to evaluate the existence of any systematics. An example is presented 
in Figure \ref{residuals} for the same index as in Figure \ref{index_sensitivity}. The average 
correlation coefficient $R^{2}$ for the fits in which the EW is the dependent variable is 
0.024 $\pm$ 0.017, with minimum and maximum values 0.006 and 0.096, respectively. For the
cases with \teff, [Fe/H] and \logg\, as the independent variables, the values of $R^{2}$ are
essentially null for all indices. These results demonstrate that the residuals are free of any 
significant systematic trends and there is no need to apply any a posteriori linear corrections 
to the EWs calculated with Equation (\ref{ewcalib}). 


\subsection{Determination of the Atmospheric Parameters}

\label{method}

The last of the three codes developed for this work delivers the final products of our 
analysis, which are the atmospheric parameters and their associated uncertainties. To 
accomplish this task, the program only requires the equivalent widths and calibrations discussed 
in sections \ref{ews} and \ref{regressions}, respectively. These can be changed in order to 
account for differences in resolution and wavelength range of the spectra. The choice of the best 
atmospheric parameters for each star is based on the minimization of a reduced chi-square 
($\chi_{r}^{2}$) and is described below. 

First, a set of 92 theoretical EWs (one for each index) is calculated for each point of a
three-dimensional grid of atmospheric parameters. The grid covers the following intervals: 
4700 K $\leq$ \teff\, $\leq$ 6600 K, with 10 K steps; $-$0.90 $\leq$ [Fe/H] $\leq$ 0.50, 
with 0.02 dex steps; 3.50 $\leq$ \logg\, $\leq$ 4.70, with 0.05 dex steps. These ranges 
are consistent with the parameter space covered by the calibration sample.  
Smaller steps were tested, but no improvements in the results were observed. Then, for 
each set of atmospheric parameters, a comparison between the observed and theoretical 
EWs is performed through the calculation of a reduced chi-square as follows:

\begin{eqnarray}
\label{redchi2}
\chi_{r}^{2} = \frac{1}{N_{ind}} \sum_{i=1}^{N_{ind}} \frac{(EW_{i}^{obs}-EW_{i}^{calc})^2}{\sigma_{i}^2}
\end{eqnarray}

In the above equation, $EW_{i}^{obs}$ and $EW_{i}^{calc}$ are the equivalent widths of 
the $i$-th index measured on the observed spectra and calculated with the calibrations, 
respectively. The number of indices used is given by $N_{ind}$. Finally, $\sigma_{i}$
is an error associated with the equivalent widths. For the calibration sample, 
it is simply the standard deviation found for each calibration, i.e., $\sigma_{i} = 
\sigma_{i}^{calib}$. For the MARVELS validation sample, $\sigma_{i}$ is given by the quadratic
sum of this term and the standard deviation of the average EWs measured on the stellar
spectra, i.e., $\sigma_{i} = \sqrt{(\sigma_{i}^{calib})^2 + (\sigma_{i}^{star})^2}$ 
(more details in section \ref{marvelsresults}).

The final atmospheric parameters are those that produce the minimum value of $\chi_{r}^{2}$.
This value is also used for the determination of the associated uncertainties. First, we
consider all sets of parameters that have $(\chi_{r}^{2})_{min} \leq \chi_{r}^{2} \leq 
2(\chi_{r}^{2})_{min}$, then calculate the differences between these parameters and those 
considered to be the best ones (corresponding to $\chi_{r}^{2})_{min}$). Finally, the root mean 
squares of these differences are taken as the errors of the parameters.

\section{Results and Discussion}

\label{results}

The previous sections were devoted to a description of the method that was used to 
derive the atmospheric parameters based only on the equivalent widths of spectral indices. 
We now discuss the tests performed to show that this approach is capable of recovering 
the precise atmospheric parameters determined from high-resolution stellar
spectra and model atmosphere analysis.

\subsection{Application of the Method to the Calibration Sample} 

\label{calibresults}

\begin{figure}
\epsscale{0.8}
\plotone{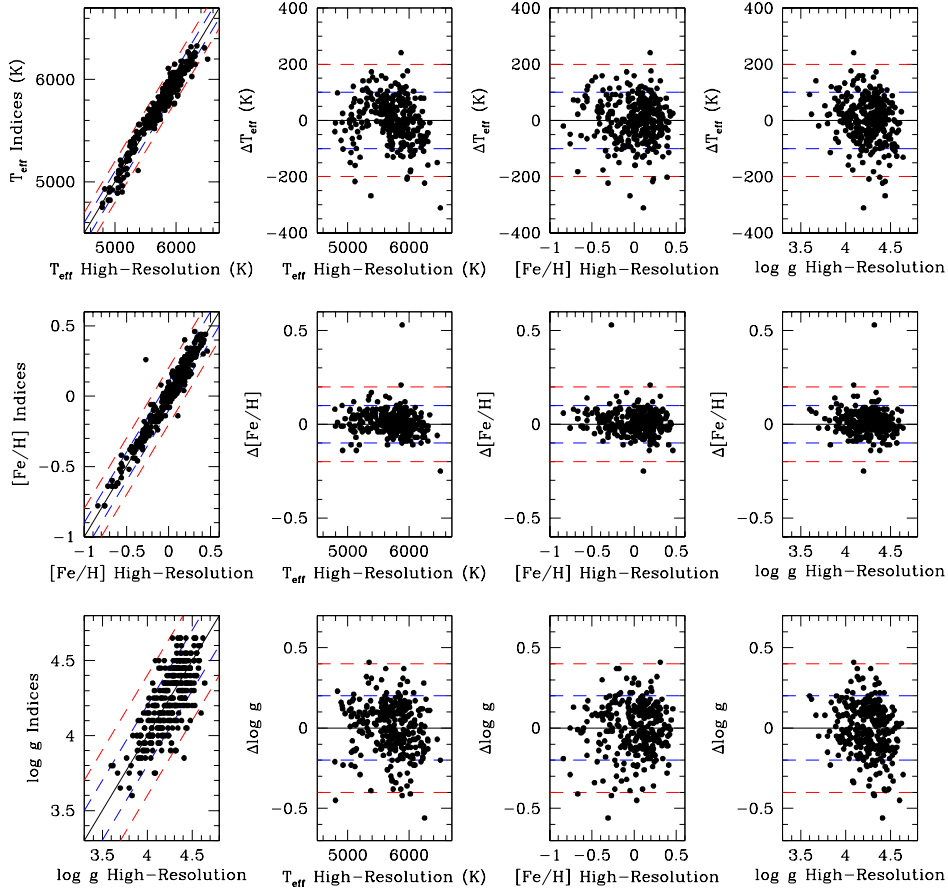}
\caption{Comparison between atmospheric parameters derived from spectral indices and 
high-resolution analyses for the calibration sample. The first column shows the 
direct comparison between the two sets of parameters, with \teff, [Fe/H] and \logg\,
on the top, middle and bottom rows, respectively. The differences between the 
results from the indices and high-resolution as a function of the latter are 
presented in the remaining panels. The differences for \teff, [Fe/H] and \logg\,
are shown in the top, middle and bottom rows, respectively. The dependency 
with the high-resolution effective temperatures, metallicities and surface gravities 
are given in columns 2, 3 and 4, respectively. In all panels, the solid black line
represents perfect agreement. The dashed blue and red lines mark, respectively, 
the offsets of $\pm$100 and $\pm$200 K for \teff, $\pm$0.10 and $\pm$0.20 dex
for [Fe/H] and $\pm$0.20 and $\pm$0.40 dex for \logg. There is a good agreement
between the two sets of atmospheric parameters and no significant offsets and trends 
in the residuals.}
\label{param_calib}
\end{figure}

The first test was a sanity check with the calibration sample. The same 
equivalent widths employed to build the calibrations were used to derive the
atmospheric parameters for the 309 calibration stars. The comparison between these
results and those derived from high-resolution analysis is presented in Figure 
\ref{param_calib}. The average differences obtained for each parameter are given
in the upper part of Table \ref{diff_param}. 

The offsets between the two sets of parameters are negligible and the dispersions
around the average residuals are lower or of the order of the external errors usually
found in high-resolution analyses (for instance, see Table 5 of \citealt{ghezzi10a}).
The situation is similar for the internal uncertainties of the method (based on 
the $\chi_{r}^{2}$; see Section \ref{method}), which have typical values in the ranges 
50 -- 150 K for \teff, 0.05 -- 0.10 dex for [Fe/H] and 0.10 -- 0.25 
dex for \logg. The obvious outlier in the middle row panels of Figure 
\ref{param_calib} is a star that was excluded for 94 indices during the regressive 
analysis (see Section \ref{regressions}).

No significant trends can be seen in the residuals, except for systematically higher 
and lower \teff\, values at $\sim$5400 K and above $\sim$6300 K, respectively. The latter 
behavior is probably caused by the reduced number of calibration stars (only seven) or by 
the fact that most of the indices become weaker in this temperature interval. This hypothesis 
is further supported by the behavior of the internal uncertainties; slight increases can 
be observed towards higher effective temperatures or lower metallicities. In both regimes, 
we have fewer stars and smaller EWs. The behavior at $\sim$5400 K could be related to the 
discontinuity in the number of calibrations stars with cooler and hotter effective temperatures 
(see Figure \ref{samples}). This systematic effect, however, has a small amplitude,
with the majority of the residuals being within $\pm$100 K. 


As a complementary check, we have also analyzed the FEROS solar spectrum from 
\cite{ghezzi10a} with the same approach as above. The derived atmospheric 
parameters were: \teff\, = 5720 $\pm$ 95 K, [Fe/H] = $-$0.02 $\pm$ 0.06 and 
\logg\, = 4.30 $\pm$ 0.16. We can see that there is a good agreement with the
standard solar parameters within the uncertainties. This result is expected 
given that the selection of the indices was based on a Ganymede 
spectrum used as solar template. The above results thus provide a confirmation 
that the spectral indices method produces internally consistent atmospheric 
parameters. 

\subsection{Application of the Method to the MARVELS Validation Sample} 

\label{marvelsresults}

The second test had the goal of demonstrating that the method is capable of accurately
recovering the high-resolution atmospheric parameters of a given stellar sample
that was not used in the calibration of the method. 
To perform this exercise, we used the validation sample of 30 stars observed as part of 
the MARVELS survey (see Section \ref{testsample}). 

The approach described in Section \ref{pipeline} was used 
to obtain the atmospheric parameters from the reduced, defringed, 1D, 
Doppler-corrected MARVELS spectra. A few remarks are needed, however. All MARVELS stars
have two sets of spectra because their light is divided into two ``beams'' that are 
collected by two neighboring fibers (see Section \ref{testsample}). Each set of 
spectra was analyzed separately and the exposures within each group were not combined 
for the normalization procedure. 
This procedure was done because we wanted to have multiple measurements for the EWs 
of the indices in order to remove possible outliers and to have an estimate of their 
uncertainties. The average EWs obtained after an iterative 2$\sigma$ clipping 
(until there were no outliers left) were used as input to the atmospheric 
parameters determination method.
The $\sigma$ used in the calculation of $\chi_{r}^{2}$ was given by the quadratic 
sum of $\sigma_{i}^{calib}$ and $\sigma_{i}^{star}$, as previously mentioned in 
Section \ref{method}. Finally, the two sets of parameters for each star were
combined using a simple arithmetic average, with the uncertainties being obtained
through an error propagation. 

The results in this initial attempt were not acceptable. The offsets relative to the 
high-resolution parameters and the dispersions were too large (see the 
upper part of Table \ref{diff_param}). Moreover, there were significant systematic 
trends as a function of \teff. The only possible causes for these issues were the 
two types of input data used in our method: the measured EWs or the calibrations.
However, the good results obtained from the test with the calibration sample 
ruled out the latter.
Direct comparisons between the equivalent widths measured 
in the FEROS degraded and MARVELS solar spectra revealed significant differences. 
However, these differences were not caused by problems in the normalization and did not 
present any systematic behavior as a function of $\lambda$, \teff, [Fe/H], \logg\, and the 
EWs themselves. Therefore, we decided to derive individual corrections 
for each index to place the MARVELS EWs onto the FEROS scale.

\subsubsection{Restriction of the Set of Indices and Correction of the EW Scale}

\label{ewcorr}

In order to derive the corrections, we have considered 120 MARVELS solar spectra 
(one for each fiber of the instrument) observed simultaneously on 17 November 2009. 
There is a different wavelength solution for each fiber, so 
the ranges covered vary from one spectrum to another. To avoid having indices with 
significantly fewer equivalent width measurements than others, we have decided to 
use only the wavelength range that is present in all 120 expsoures, which 
corresponds to the interval $\sim$5137 - 5543~\AA. As a consequence of this choice,
18 indices were removed from our analysis: 2 - 7 on the blue end (note that 
1 was already excluded because its calibration had $R^{2} <$ 0.9) and 85 - 96 
on the red end (all marked with a number 3 in the notes of Table 
\ref{indices_properties}). 

For the remaining 74 indices, the average EWs from the 120 solar spectra and their
respective standard deviations were calculated. No clipping was performed because 
we wanted to retain all the information regarding the variations of the EWs with
the fiber number in the instrument. This is important because the MARVELS stars could 
have been observed with any of the 120 fibers. We then calculated the 
differences $\mid\langle {\rm EW_{MARVELS}} \rangle - {\rm EW_{FEROS}}\mid$. If 
these differences were higher than the corresponding values for $\sigma(\langle 
{\rm EW_{MARVELS}} \rangle)$, the indices were removed. This was the case for 
the following ten indices (marked with a number 4 in the notes of Table 
\ref{indices_properties}): 33, 34, 39, 40, 54, 55, 61, 62, 66, 67. 

Finally, it was decided that the remaining 64 indices would have their average 
MARVELS equivalent widths corrected by the difference $\mid\langle {\rm EW_{MARVELS}} 
\rangle - {\rm EW_{FEROS}}\mid$. Although this correction was based only on the Sun,
it was applied to all stars in the validation sample, regardless of their parameters. 
As it is shown below, the correction proved to be effective. Another
constraint implemented for the analysis of the MARVELS stars was the removal of 
corrected average EWs that were outside the ranges defined by each of the calibrations 
(see Table \ref{calibrations_ew}). This cut was done to avoid any extrapolation 
outside the parameter space defined by our calibration sample.    

\subsubsection{Final Results}

\label{finalresults}

Using the method with the above optimization to the MARVELS data, we rederived the
atmospheric parameters for the 30 stars in the MARVELS validation sample. Recall that 
the final results for each object were obtained from an arithmetic average of the 
parameters produced by its two associated fibers. The two sets of results 
are in good agreement, with average differences of $-$18 $\pm$ 91 K for \teff, 0.01 $\pm$ 
0.06 dex for [Fe/H] and $-$0.09 $\pm$ 0.16 dex for \logg. These values highlight that the 
data from adjacent fibers on the MARVELS instrument are consistent.

\begin{figure}
\epsscale{1.0}
\plotone{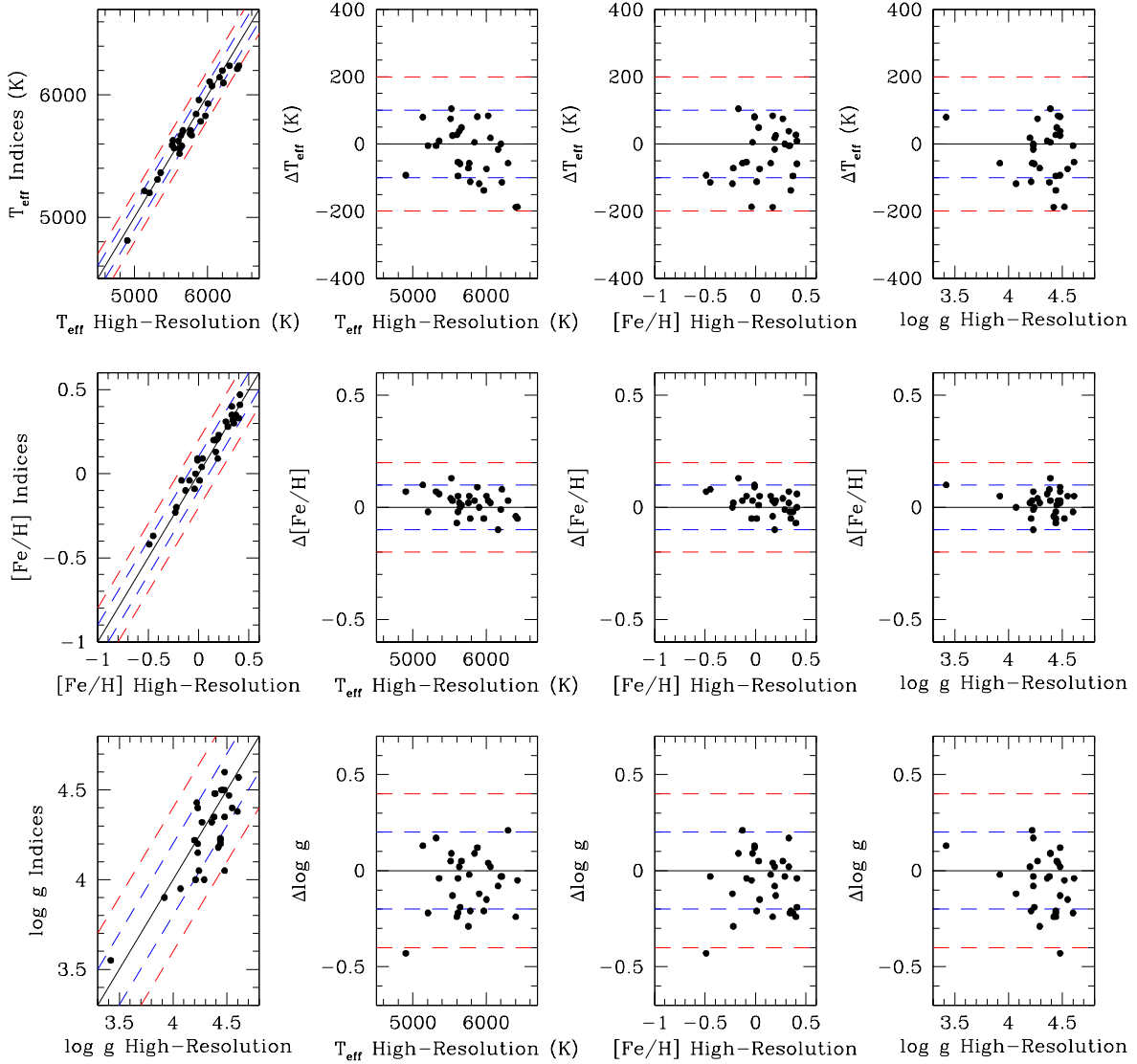}
\caption{Comparison between atmospheric parameters derived from spectral indices and 
high-resolution analyses for the MARVELS validation sample. The panels, symbols and 
lines have the same meaning as in Figure \ref{param_calib}. A good agreement between 
the two sets of atmospheric parameters can be observed and there are no clear offsets 
and trends in the residuals.}
\label{param_marvels}
\end{figure}

The comparison between the atmospheric parameters derived from the indices 
and high-resolution analyses is presented in Figure \ref{param_marvels}. The average 
differences obtained for each parameter are given in the lower part of Table 
\ref{diff_param}. The offset in the metallicity is once again close to zero. 
For \teff, we have a non-negligible
negative value, but it is lower than the dispersion of the residuals and 
the typical uncertainties found for this parameter. A similar situation was found 
for \logg. However, the offset in this case can be traced back to the reference values
for the surface gravity. We recall that the values for the calibration sample present
a small offset relative to other literature values (see Table 5 of \citealt{ghezzi10a}). 


The dispersions around the average offsets are either lower than, or on the order of,  
the typical external uncertainties of high-resolution analyses. This was 
also the case for the internal uncertainties of the method, which have typical values 
in the ranges 80 -- 160 K for \teff, 0.05 -- 0.10 dex for [Fe/H] and 0.15 -- 0.25 
dex for \logg. All these values are close to the ones derived for the calibration sample. 
The residuals do not exhibit any significant trends, except possibly for the same ones 
observed for \teff\, in the results for the calibration sample. 

\begin{figure}
\epsscale{0.9}
\plotone{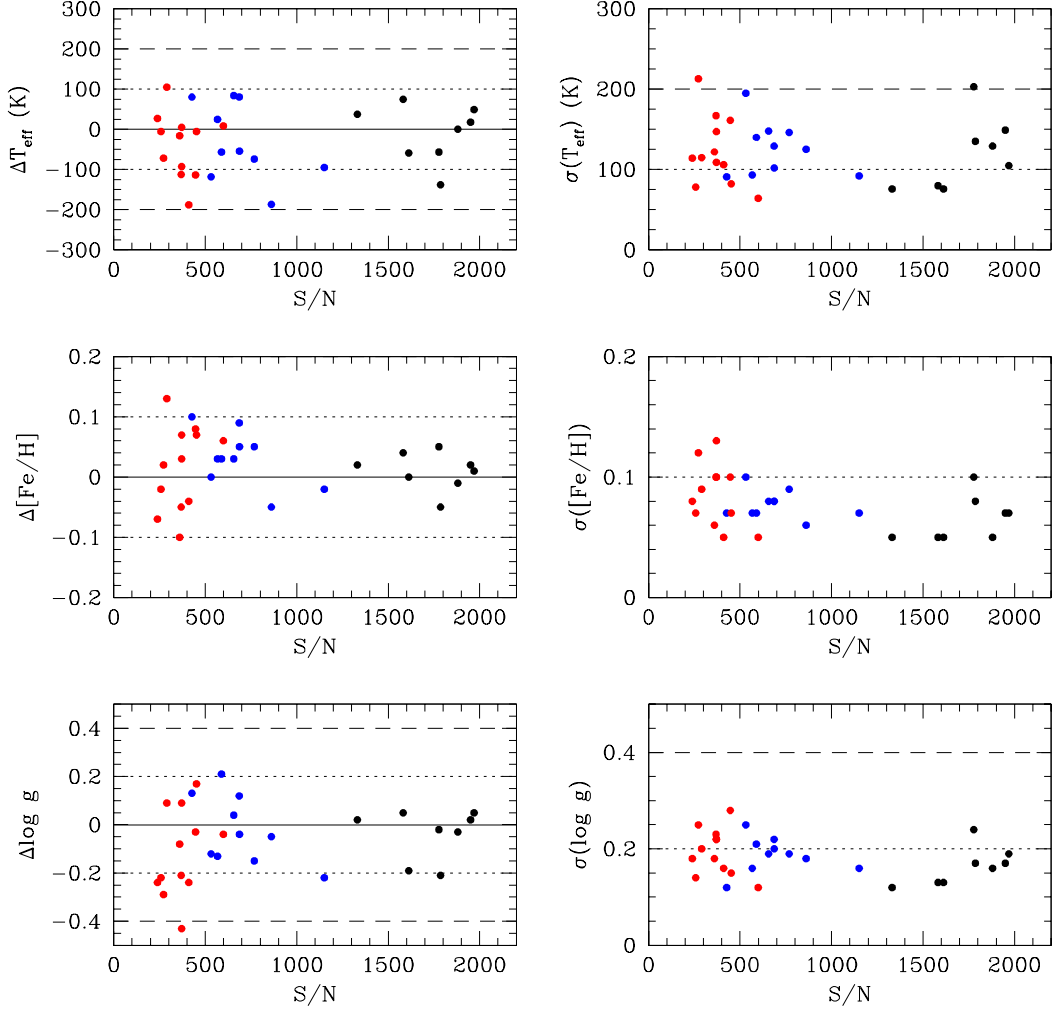}
\caption{\textit{Left column:} Differences between the atmospheric parameters derived 
from spectral indices and high-resolution analysis for the MARVELS validation sample as a 
function of the total S/N. \textit{Right column:} Internal uncertainties of our method
as a function of the total S/N. The upper, middle and lower panels show the cases for
\teff, [Fe/H] and \logg, respectively. Stars with $V$ $<$ 9, 9 $\leq$ $V$ $<$ 11 and $V$ 
$\geq$ 11 are shown by black, blue and red filled circles, respectively.
In all panels, the solid black line represents perfect agreement. The dotted and dashed 
black lines mark, respectively, the limits $\pm$100 and $\pm$200 K for \teff, $\pm$0.10 
and $\pm$0.20 dex for [Fe/H] and $\pm$0.20 and $\pm$0.40 dex for \logg. The differences 
between the two sets of atmospheric parameters and the internal uncertainties of the 
spectral indices method do not show any dependencies with S/N and $V$ magnitude.}
\label{delta_sig_sn}
\end{figure}

The residuals and internal uncertainties do not exhibit significant trends as a 
function of the S/N values of the spectra or the $V$ magnitudes of the stars (taken 
from the Guide Star Catalog; \citealt{lasker08}), as can be seen on Figure 
\ref{delta_sig_sn}. The total S/N for a given star was obtained from the 
arithmetic average of the values derived from its two sets of spectra (one from each
fiber). The S/N for each set was calculated by multiplying the average 
S/N of the individual spectra $\langle (S/N)_{spec} \rangle$ by the square root of 
the number of spectra $\sqrt{N_{spec}}$. This is a reasonable approximation considering
that the exposure times are similar for all spectra of a given star. 

The lack of a dependency between the accuracy and precision (differences and errors in left 
and right columns of Figure \ref{delta_sig_sn}, respectively) of the results and the S/N 
of the spectra can be explained by a particular characteristic of our data. All stars 
have S/N $\geq$ 200 per pixel (dispersion 0.154~\AA/pixel) and this value is certainly 
higher than the limit below which
we would start to observe a decline in the quality of the results with decreasing S/N. 
Thus, we conclude that all stars have high-quality data that allow our approach 
to work at its best precision and accuracy. 


The method was further tested using the 120 MARVELS solar spectra described previously.
Since we wanted to obtain the best possible results, an iterative 2$\sigma$ 
clipping (until there were no outliers left) was applied to calculate the average EWs
of the Sun. Following the exact same method employed for
the MARVELS validation sample, we derived the results: \teff\, = 5760 
$\pm$ 81 K, [Fe/H] = $-$0.04 $\pm$ 0.05 and \logg\, = 4.40 $\pm$ 0.15. The agreement
with the standard solar parameters is excellent. 

For completeness, we also utilized the restricted set of 64 indices (without 
any corrections) to rederive the parameters for the calibration stars.
As can be seen on the lower part of Table \ref{diff_param}, there are no significant 
differences relative to the previous case in which all 92 indices were considered. 

The above results for the MARVELS validation sample show that the spectral indices
approach is able to accurately recover the precise atmospheric parameters 
derived from the analysis of high-resolution spectra. Moreover, the differences 
between the two sets of results and the internal uncertainties of the method 
presented here are both consistent with the typical external errors obtained in 
high-resolution analyzes. It should be highlighted that this is achieved by using 
only equivalent widths directly measured on the observed low-resolution MARVELS 
spectra, with no other priors. 

\subsection{Application of the Method to the ELODIE Stellar Library} 

\label{elodieresults}

The third and last test was conducted to complement the previous one. Although the 
validation sample contains stars with real spectra from the MARVELS survey, we recognize
that 30 stars is a relatively small number to critically evaluate the performance of our
method. Therefore, we decided to analyze the spectra contained in the ELODIE stellar library
(\citealt{ps01}, \citealt{ps07}; see website\footnote{http://www.obs.u-bordeaux1.fr/m2a/soubiran/elodie\_library.html} 
for the most updated 3.1 version). 
Briefly, the library includes 1962 high-resolution (R = 42,000)
spectra observed with ELODIE spectrograph coupled to the 1.93 m telescope at Observatoire de 
Haute-Provence. The spectra cover the wavelength range 3900 - 6800 \AA, have S/N values varying
between 30 and 680 and belong to 1388 stars that non-uniformly sample the following intervals of 
atmospheric parameters: 3442 K $\leq$ \teff\, $\leq$ 47250 K, $-$2.94 $\leq$ [Fe/H] 
$\leq$ $+$1.40 and 0.00 $\leq$ \logg\, $\leq$ 4.90.

The above values are quoted from the source catalog for reference, but our analysis is restricted to
only those stars who have literature average atmospheric 
parameters within the limits defined by our calibration sample (4800 K $\lesssim$ \teff\, $\lesssim$ 
6500 K, $-$0.90 $\lesssim$ [Fe/H] $\lesssim$ $+$0.50 and 3.60 $\lesssim$ \logg\, $\lesssim$ 4.70).
We have also restricted our selection to stars with classes 3 or 4 attributed to their average 
parameters, which means that the typical standard deviations around their mean \teff\, and [Fe/H]
values are 62 K and 0.08 dex and 74 K and 0.10 dex, respectively\footnote{For more details, 
check \url{http://www.obs.u-bordeaux1.fr/m2a/soubiran/elo\_stel\_param.html}}. No cuts in S/N were applied,
but stars with indications of variability or close neighbors were removed. The final ELODIE test 
sample consists of 219 spectra from 138 unique stars. As for the FEROS sample, these spectra were
degraded to the MARVELS resolution following the same procedure described in Section \ref{calibsample}.

\begin{figure}
\epsscale{1.0}
\plotone{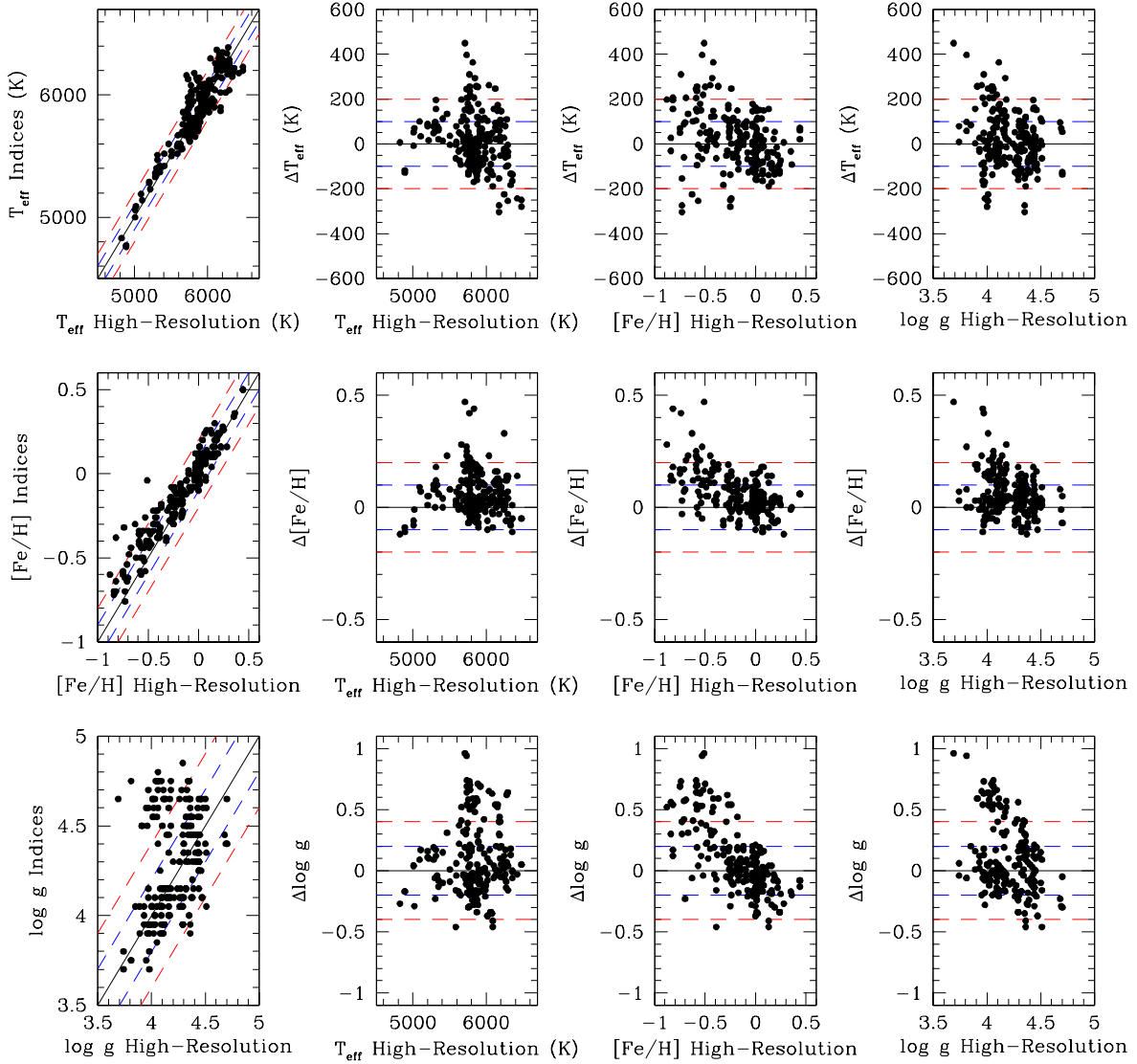}
\caption{Comparison between atmospheric parameters from spectral indices and 
the literature for the ELODIE test sample. The panels, symbols and 
lines have the same meaning as in Figure \ref{param_calib}. Note, however, that the scales 
are different in some panels. A good agreement between the two sets of atmospheric parameters 
can be observed and small offsets and trends in the residuals.}
\label{param_elodie}
\end{figure}

The ELODIE test sample was analyzed with the approach presented in Section \ref{pipeline}.
All spectra were normalized individually and the equivalent widths of the indices
were measured in each of them, as if they belonged to different stars. This was done to check the internal 
consistency of the atmospheric parameter determinations for a given star based on different spectra,
which found to be good. The EWs of the 92 indices for which the calibrations have R$^{2} >$ 0.9 were 
considered. Again, a data-specific tuning was necessary and a correction to the equivalent width scale 
based on the Sun was performed. Average solar ELODIE equivalent 
widths were calculated using 11 of the 13 spectra of the Sun with S/N $>$ 100. The differences between these
average EWs and the values measured on the FEROS solar degraded spectrum were adopted as individual 
corrections for each index. Using these corrected equivalent widths and the calibrations presented in 
Section \ref{regressions}, atmospheric parameters were derived for the ELODIE sample. The results can be
seen in Figure \ref{param_elodie} and in Table \ref{diff_param}.


\begin{figure}
\epsscale{1.0}
\plotone{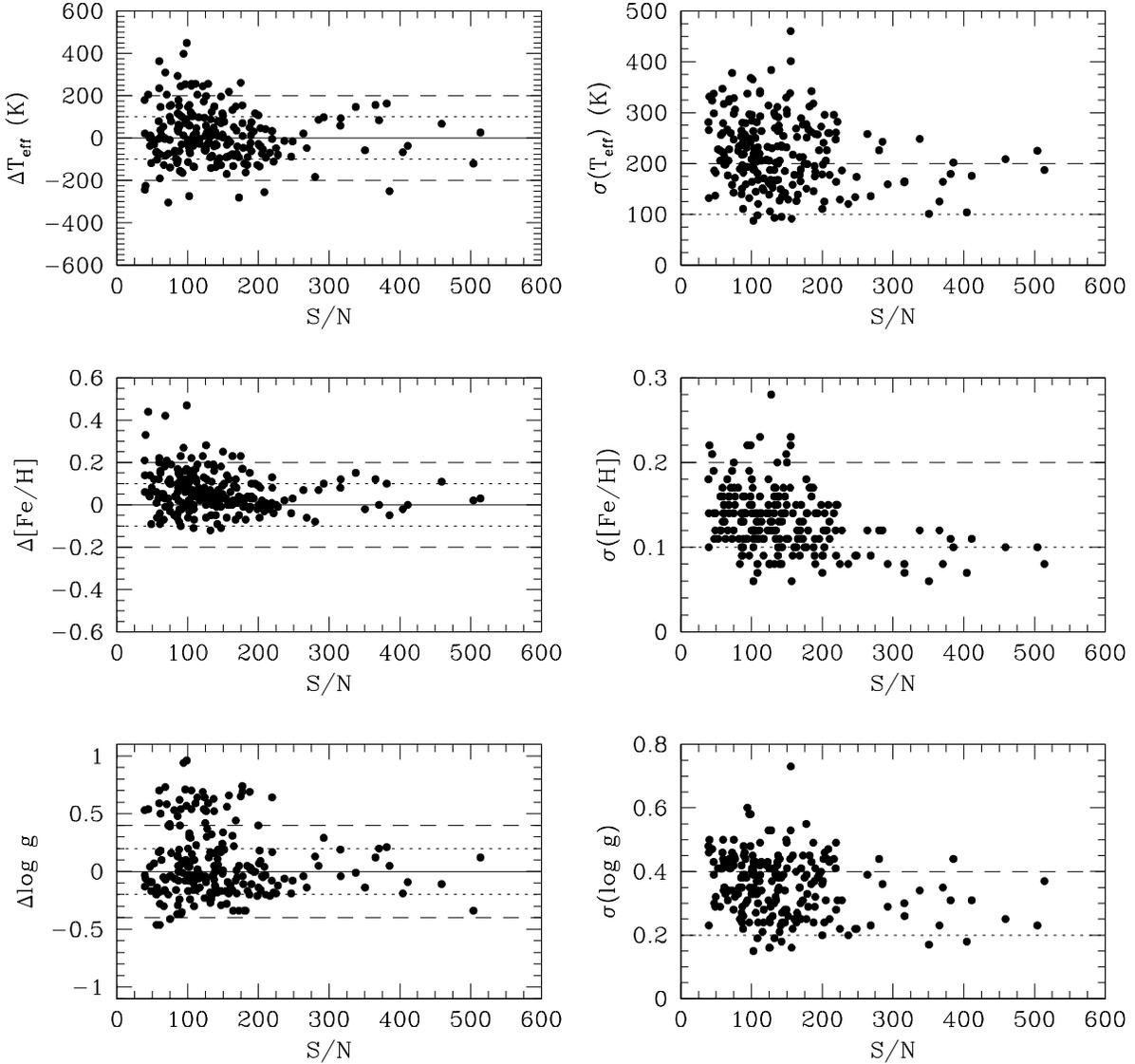}
\caption{\textit{Left column:} Differences between the atmospheric parameters  
from spectral indices and from the literature for the ELODIE test sample as a 
function of S/N per pixel. \textit{Right column:} Internal uncertainties of our method
as a function of S/N per pixel. The panels, symbols and lines have the same meaning as in 
Figure \ref{delta_sig_sn}. Note, however, that the scales are different and no distinctions
based on V magnitude are made. The differences between the two sets of atmospheric parameters 
and the internal uncertainties of the spectral indices method significantly increase for 
S/N $<$ 200 - 250.}
\label{delta_sig_sn_elodie}
\end{figure}

We can see that the agreement for \teff\, and [Fe/H] is good and the average differences 
and dispersions (15 $\pm$ 125 K and 0.06 $\pm$ 0.10,
respectively) can be entirely explained by a combination of the internal and external uncertainties
from our method with the typical standard deviations around the mean ELODIE parameters. Mild trends
are observed for the residuals of these parameters, with the differences between the two sets of 
results increasing for hot and metal-poor stars,
which is expected due to the lower number of calibration stars in these regions of the parameter
space. Spectra with low S/N values also contribute for the observed trend, since they are clearly
associated with the largest residuals (see also Figure \ref{delta_sig_sn_elodie}).

The agreement for \logg, on the other hand, is only reasonable, with a relatively large dispersion
and clear trends in the residuals. This result was not totally unexpected since
the previous tests with the calibration and validation samples have shown that the surface gravity is 
the most difficult parameter to accurately constrain. Although the average difference and dispersion 
(0.07 $\pm$ 0.29) can again be explained by a combination of the uncertainties in the
two sets of parameters, we can see that some stars have differences higher than 0.5 dex. However,
it is clear from the lower panels that these stars are hot and/or metal-poor. Thus, the poor 
determinations of \teff\, and [Fe/H] could also be affecting the estimate of \logg. Many of these stars
also have spectra with low S/N values (see Figure \ref{delta_sig_sn_elodie}).

The average solar parameters derived from the 13 spectra available are: \teff\, = 5750 $\pm$ 66 K,
[Fe/H] = 0.00 $\pm$ 0.04 and \logg\, = 4.32 $\pm$ 0.11. The agreement with the canonical solar 
parameters \teff\, = 5777 K, [Fe/H] = 0.00 and \logg\, = 4.44) is very good, with the surface 
gravity providing the most discrepant result once again.

The ELODIE test sample also offers the opportunity to check the performance of our method for lower
S/N spectra, which was not possible for the MARVELS validation sample. Figure \ref{delta_sig_sn_elodie}
shows the differences between the spectral indices and ELODIE atmospheric parameters as well
as the uncertainties of the former as a function of the S/N per pixel of the spectra. For S/N $>$ 200 
- 250, the typical values of the differences and uncertainties agree well with the ones for the 
MARVELS sample. At S/N $\sim$ 200 - 250, there is a significant increase in both quantities. 
For S/N $<$ 200 - 250, the scatter of the distributions stay pretty much constant. These results 
show that our method is capable of providing reliable results for spectra with lower S/N, although
with lower accuracies than in the high-S/N regime. 


The tests discussed above provide an additional validation of the spectral indices
method, based on a larger sample of observed stellar spectra with typically lower S/N values. 
As in the previous sections, the precise atmospheric parameters from the literature were
recovered with a good precision. The small trends observed in the residuals
are caused by the non-uniform sampling of the parameter space by our calibration sample
and the lower sensitivities of the indices to the surface gravity.

\section{Conclusions}

\label{conclusions}

In this paper, we described the usage of spectral indices as a fast and homogeneous 
approach to determine accurate atmospheric parameters (\teff, [Fe/H], \logg) for 
samples of FGK dwarfs and subgiants with intermediate (R $\sim$ 12,000) resolution 
spectra, applied to observations from the MARVELS survey. Equivalent widths of the indices
were measured on normalized spectra and then compared to values calculated with
a set of calibrations, which were constructed using 309 stars with precise atmospheric
parameters resulting from the analysis of FEROS spectra. The best solutions were
obtained through a $\chi_{r}^{2}$ minimization of the differences between these 
two sets of EWs. The entire analysis was automatized with three codes developed
for this work.

The spectral indices method was validated with a sample of 30 stars which were 
observed as part of the MARVELS survey and have precise atmospheric parameters 
derived from classical model atmosphere analyses. We were able to 
recover these parameters within 80 K for \teff, 0.05 dex for [Fe/H] and 0.15 dex 
for \logg. The average internal errors of the method are 101 K, 0.06 dex and 0.15 
dex, respectively. These values are consistent with the typical external 
uncertainties found between different high-resolution spectroscopic analyses 
(see, e.g., Table 5 of \citealt{ghezzi10a}).

Further confirmation of the good performance of the method was provided by
the analysis of a subsample of 138 stars from the ELODIE stellar library
with average literature atmospheric parameters with good quality. These were
recovered by our method with precisions of 125 K for \teff, 0.10 dex for [Fe/H] 
and 0.29 dex for \logg. The higher uncertainties are caused by a combination
of the internal uncertainties of the spectral indices approach, the typically lower
S/N values for the ELODIE spectra and the typical
standard deviations around the mean ELODIE parameters. Undersampled regions
in the parameter space for the calibration sample might also play a role.
However, these errors are still similar or lower than typical
uncertainties provided by the pipelines of other intermediate-resolution
surveys (e.g., RAVE; \citealt{kordopatis13}).

This agreement demonstrates that the spectral indices approach is a powerful
tool to derive accurate and precise atmospheric parameters for solar-type stars. 
The method is quite general and its application to a particular data set requires 
only a small tuning of the input parameters (list of indices and the measured 
equivalent widths). The approach has an excellent performance, yielding typical stellar
parameter uncertainties comparable to or better than those provided by
pipelines that utilize spectral synthesis, such as the ones
from the SEGUE, RAVE and LAMOST surveys. We note, however, that they analyze
spectra with lower resolutions, S/N values, wavelength coverages or combinations of these
properties. Our method could be adopted by similar surveys, with modifications to 
the list of indices and calibrations as appropriate for differences in the resolution and 
wavelength coverage.
 
The next step of our work is the application of the spectral indices method to the 
entire MARVELS sample ($\sim$3,300 stars), which contains many faint stars ($V$ 
$\sim$ 10-12) that were never previously analyzed. The atmospheric parameters will then 
be used to derive additional stellar properties, such as mass, radius, distances and 
ages. The kinematic analysis, on the other hand, will follow from the precise RVs 
determined by MARVELS. With accurate chemical and kinematical information for a 
large and statistically homogeneous sample, MARVELS can provide valuable contributions
to many studies, such as the comparison of the statistical properties 
of stars with and without companions, the search for correlations between the properties 
of the companions and their stellar hosts, and Galactic chemical and dynamical evolution 
in the solar neighborhood. These will be the subjects of subsequent papers. 


\acknowledgments

Funding for SDSS-III has been provided by the Alfred P. Sloan Foundation, the Participating 
Institutions, the National Science Foundation, and the U.S. Department of Energy Office of 
Science. The SDSS-III web site is \url{http://www.sdss3.org/}.

SDSS-III is managed by the Astrophysical Research Consortium for the Participating 
Institutions of the SDSS-III Collaboration including the University of Arizona, the Brazilian 
Participation Group, Brookhaven National Laboratory, Carnegie Mellon University, University 
of Florida, the French Participation Group, the German Participation Group, Harvard University, 
the Instituto de Astrofisica de Canarias, the Michigan State/Notre Dame/JINA Participation 
Group, Johns Hopkins University, Lawrence Berkeley National Laboratory, Max Planck Institute 
for Astrophysics, Max Planck Institute for Extraterrestrial Physics, New Mexico State 
University, New York University, Ohio State University, Pennsylvania State University, 
University of Portsmouth, Princeton University, the Spanish Participation Group, University 
of Tokyo, University of Utah, Vanderbilt University, University of Virginia, University of 
Washington, and Yale University.

Funding for the Brazilian Participation Group has been provided by the 
Minist\'erio de Ci\^encia e Tecnologia (MCT), Funda\c c\~ao Carlos Chagas 
Filho de Amparo \`a Pesquisa do Estado do Rio de Janeiro (FAPERJ), Conselho 
Nacional de Desenvolvimento Cient\'ifico e Tecnol\'ogico (CNPq), and 
Financiadora de Estudos e Projetos (FINEP).

We thank Eduardo del Peloso for developing the codes used to normalize
the spectra and measure the equivalent widths of the indices.
We acknowledge Katia Cunha, Verne Smith and Daniel Eisenstein for helpful
suggestions. We thank the anonymous referee for the thorough reading of 
the manuscript and the detailed comments which helped improving it.

L.G. acknowledges financial support provided by the PAPDRJ CAPES/FAPERJ 
Fellowship. L.G. thanks K. Cunha and V. Smith for discussions and for
helping with the acquisition of the FEROS spectra.
L.D.F. and D.L.O. acknowledge financial support from CAPES and ESO student fellowships. 
G.F.P.M. acknowledges the financial support by CNPq (476909/2006-6 and
474972/2009-7) and FAPERJ (APQ1/26/170.687/2004) grants.
B.X.S. acknowledges support from CNPq (301462/2009-7).



{\it Facilities:} \facility{Max Planck:2.2m}, \facility{Sloan}, \facility{ARC}



\begin{deluxetable}{lcrcc}
\tablecolumns{5}
\tablewidth{0pt}
\tablecaption{Atmospheric Parameters for the MARVELS Candidates.
\label{marvels_cand_param}}
\tablehead{ \colhead{Star} & \colhead{\teff} & \colhead{[Fe/H]} 
& \colhead{\logg} & \colhead{Reference} \\ 
\colhead{} & \colhead{(K)} & \colhead{} & \colhead{} & \colhead{} } 
\startdata
MC1        & 6297 $\pm$ 28 & $-$0.13 $\pm$ 0.04 & 4.22 $\pm$ 0.09 & \cite{wright13} \\
MC2        & 5598 $\pm$ 63 &    0.40 $\pm$ 0.09 & 4.44 $\pm$ 0.17 & \cite{delee13} \\
MC5        & 6214 $\pm$ 38 & $-$0.45 $\pm$ 0.06 & 4.38 $\pm$ 0.15 & This work \\
MC6        & 6427 $\pm$ 33 & $-$0.04 $\pm$ 0.05 & 4.52 $\pm$ 0.14 & \cite{fleming12} \\
MC7        & 5879 $\pm$ 29 & $-$0.01 $\pm$ 0.05 & 4.48 $\pm$ 0.15 & \cite{wisniewski12} \\
MC11       & 6004 $\pm$ 29 &    0.04 $\pm$ 0.05 & 4.55 $\pm$ 0.15 & \cite{jiang13} \\
MC12       & 5903 $\pm$ 42 & $-$0.23 $\pm$ 0.07 & 4.07 $\pm$ 0.16 & \cite{ma13} \\
MCKGS1-50  & 5540 $\pm$ 39 &    0.20 $\pm$ 0.07 & 4.48 $\pm$ 0.23 & This work \\
MCKGS1-52  & 6403 $\pm$ 44 &    0.17 $\pm$ 0.07 & 4.42 $\pm$ 0.24 & This work \\
MCKGS1-61  & 5757 $\pm$ 43 & $-$0.22 $\pm$ 0.07 & 4.29 $\pm$ 0.22 & This work \\
MCKGS1-70  & 5135 $\pm$ 39 & $-$0.01 $\pm$ 0.08 & 3.42 $\pm$ 0.18 & This work \\
MCKGS1-94  & 4903 $\pm$ 47 & $-$0.49 $\pm$ 0.08 & 4.48 $\pm$ 0.28 & This work \\
MCKGS1-112 & 5782 $\pm$ 42 &    0.01 $\pm$ 0.07 & 4.21 $\pm$ 0.23 & This work \\
MCKGS1-135 & 5525 $\pm$ 71 & $-$0.17 $\pm$ 0.08 & 4.39 $\pm$ 0.23 & This work \\
MCKGS1-153 & 5614 $\pm$ 29 & $-$0.09 $\pm$ 0.06 & 4.61 $\pm$ 0.09 & This work \\
MCUF1-11   & 5315 $\pm$ 44 &    0.33 $\pm$ 0.06 & 4.23 $\pm$ 0.19 & This work \\
\enddata
\end{deluxetable}


\begin{deluxetable}{llrcc}
\tablecolumns{5}
\tablewidth{0pt}
\tablecaption{Atmospheric Parameters for the MARVELS Reference Stars.
\label{marvels_ref_param}}
\tablehead{ \colhead{Star} & \colhead{\teff} & \colhead{[Fe/H]} 
& \colhead{\logg} & \colhead{References\tablenotemark{a}} \\ 
\colhead{} & \colhead{(K)} & \colhead{} & \colhead{} & \colhead{} } 
\startdata
WASP 1    & 6161 $\pm$ 52 &     0.19 $\pm$ 0.06 & 4.23 $\pm$ 0.05 & 1,2,3 \\
HD 4203   & 5644 $\pm$ 61 &     0.41 $\pm$ 0.03 & 4.24 $\pm$ 0.14 & 4,5,6,7,8,9,10,11 \\
HD 9407   & 5661 $\pm$  5 &     0.03 $\pm$ 0.02 & 4.45 $\pm$ 0.03 & 12,8,13 \\
HD 17156  & 6057 $\pm$ 46 &     0.19 $\pm$ 0.05 & 4.20 $\pm$ 0.11 & 1,14,7,15,16,17 \\
HIP 14810 & 5515 $\pm$ 25 &     0.27 $\pm$ 0.02 & 4.27 $\pm$ 0.06 & 18,14,5,7,19 \\
HD 43691  & 6200 $\pm$ 39 &     0.29 $\pm$ 0.02 & 4.23 $\pm$ 0.12 & 14,18,7,20 \\
HD 49674  & 5632 $\pm$ 31 &     0.33 $\pm$ 0.01 & 4.48 $\pm$ 0.12 & 18,7,8,9 \\
XO-2      & 5356 $\pm$ 19 &     0.41 $\pm$ 0.05 & 4.36 $\pm$ 0.19 & 1,16,21 \\
HD 68988  & 5968 $\pm$ 48 &     0.35 $\pm$ 0.02 & 4.44 $\pm$ 0.08 & 18,7,22,8,9,11,23 \\
HD 80606  & 5615 $\pm$ 70 &     0.37 $\pm$ 0.08 & 4.44 $\pm$ 0.07 & 1,7,22,8,9,23,10,24 \\
HD 118203 & 5767 $\pm$ 70 &     0.15 $\pm$ 0.06 & 3.92 $\pm$ 0.04 & 25,18,7 \\
HAT-P-3   & 5205 $\pm$ 28 &     0.34 $\pm$ 0.10 & 4.60 $\pm$ 0.02 & 1,26 \\
TReS-2    & 5840 $\pm$ 41 &  $-$0.03 $\pm$ 0.11 & 4.39 $\pm$ 0.09 & 1,16,27 \\
HAT-P-1   & 6026 $\pm$ 71 &     0.17 $\pm$ 0.06 & 4.46 $\pm$ 0.01 & 16,28 \\
\enddata
\tablenotetext{a}{The references for each star are organized in 
reverse chronological order.}
\tablerefs{
(1) \cite{torres12};
(2) \cite{albrecht11};
(3) \cite{stempels07};
(4) \cite{saffe11};
(5) \cite{ghezzi10a};
(6) \cite{gh10};
(7) \cite{gonzalez10};
(8) \cite{vf05};
(9) \cite{santos04};
(10) \cite{santos03};
(11) \cite{laws03};
(12) \cite{mishenina08};
(13) \cite{mg00};
(14) \cite{kang11};
(15) \cite{barbieri09};
(16) \cite{AvE09};
(17) \cite{fischer07};
(18) \cite{brugamyer11};
(19) \cite{wright09};
(20) \cite{dasilva07};
(21) \cite{burke07};
(22) \cite{lh06};
(23) \cite{hl03};
(24) \cite{naef01};
(25) \cite{zielinski12};
(26) \cite{torres07};
(27) \cite{sozzetti07};
(28) \cite{bakos07}.}
\end{deluxetable}


\begin{deluxetable}{ccccc}
\tablecolumns{5}
\tablewidth{0pt}
\tablecaption{Properties of the Spectral Indices.
\label{indices_properties}}
\tablehead{ \colhead{Index} & \colhead{$\lambda_{i}$} & \colhead{$\lambda_{f}$} 
& \colhead{Dominant Lines} & \colhead{Notes} \\ 
\colhead{} & \colhead{(\AA)} & \colhead{(\AA)} & \colhead{} & \colhead{}} 
\startdata
1  & 5100.40 & 5102.00 & Fe II                    & 2 \\ 
2  & 5102.00 & 5104.90 & Ni I, Fe I               & 3 \\ 
3  & 5105.10 & 5106.05 & Cu I                     & 3 \\ 
4  & 5106.90 & 5108.25 & Fe I                     & 3 \\ 
5  & 5109.95 & 5111.05 & Fe I                     & 3 \\ 
6  & 5130.95 & 5132.20 & Fe I, Ni I               & 3 \\ 
7  & 5135.55 & 5137.90 & Fe I, Ni I               & 3 \\ 
8  & 5138.05 & 5140.45 & Fe I, Cr I               & 1 \\ 
9  & 5140.65 & 5144.15 & Fe I, Ni I               & 1 \\ 
10 & 5194.35 & 5196.95 & Fe I, Cr I, Mn I         & 1 \\ 
11 & 5196.95 & 5198.00 & Fe II, Fe I, Ni I        & 1 \\ 
12 & 5198.25 & 5199.35 & Fe I                     & 1 \\ 
13 & 5201.55 & 5203.30 & Fe I                     & 1 \\ 
14 & 5203.30 & 5209.45 & Cr I, Fe I               & 1 \\ 
15 & 5213.55 & 5216.90 & Fe I                     & 1 \\ 
16 & 5229.10 & 5230.95 & Fe I                     & 1 \\ 
17 & 5231.85 & 5233.70 & Fe I                     & 1 \\ 
18 & 5234.05 & 5236.75 & Fe II, Fe I              & 1 \\ 
19 & 5236.75 & 5238.00 & Cr II                    & 1 \\ 
20 & 5241.25 & 5244.85 & Fe I                     & 1 \\ 
21 & 5249.75 & 5254.25 & Fe I                     & 1 \\ 
22 & 5254.25 & 5256.35 & Cr I, Fe I, Mn I         & 1 \\ 
23 & 5272.60 & 5274.05 & Fe I                     & 1 \\ 
24 & 5274.05 & 5276.80 & Fe II, Fe I, Cr I, Cr II & 1 \\ 
25 & 5279.35 & 5280.95 & Cr I, Fe I, Cr II        & 1 \\ 
26 & 5280.95 & 5282.85 & Fe I                     & 1 \\ 
27 & 5282.85 & 5286.65 & Fe I, Ti I, Fe II        & 1 \\ 
28 & 5288.00 & 5289.10 & Fe I                     & 1 \\ 
29 & 5296.15 & 5299.45 & Cr I, Fe I               & 1 \\ 
30 & 5301.50 & 5303.25 & Fe I                     & 1 \\ 
31 & 5306.75 & 5308.00 & Fe I                     & 1 \\ 
32 & 5308.00 & 5309.45 & Cr II                    & 1 \\ 
33 & 5312.10 & 5314.25 & Cr II, Cr I              & 4 \\ 
34 & 5314.25 & 5315.65 & Fe I                     & 4 \\ 
35 & 5315.90 & 5317.90 & Fe II                    & 1 \\ 
36 & 5320.50 & 5322.65 & Fe I                     & 1 \\ 
37 & 5322.65 & 5325.00 & Fe I                     & 1 \\ 
38 & 5327.05 & 5330.85 & Fe I, Cr I               & 1 \\ 
39 & 5332.05 & 5333.55 & Fe I, V II               & 4 \\ 
40 & 5334.55 & 5335.60 & Cr II, Co I              & 4 \\ 
41 & 5336.30 & 5337.35 & Ti II                    & 2 \\ 
42 & 5338.90 & 5341.95 & Fe I, Cr I, Mn I         & 1 \\ 
43 & 5341.95 & 5344.05 & Co I, Fe I               & 1 \\ 
44 & 5345.10 & 5346.40 & Cr I                     & 1 \\ 
45 & 5347.35 & 5348.90 & Cr I                     & 1 \\ 
46 & 5352.65 & 5354.20 & Fe I, Co I, Ni I         & 1 \\ 
47 & 5360.70 & 5362.15 & Fe I                     & 1 \\ 
48 & 5362.20 & 5363.65 & Fe II, Fe I, Co I        & 1 \\ 
49 & 5363.65 & 5366.05 & Fe I                     & 1 \\ 
50 & 5366.90 & 5368.15 & Fe I                     & 1 \\ 
51 & 5368.95 & 5372.85 & Fe I, Ni I, Co I         & 1 \\ 
52 & 5373.10 & 5374.70 & Fe I                     & 1 \\ 
53 & 5377.10 & 5378.55 & Mn I                     & 1 \\ 
54 & 5382.65 & 5384.35 & Fe I                     & 4 \\ 
55 & 5385.85 & 5388.00 & Fe I, Cr I               & 4 \\ 
56 & 5388.85 & 5392.30 & Fe I                     & 1 \\ 
57 & 5392.50 & 5394.05 & Fe I                     & 1 \\ 
58 & 5394.05 & 5395.75 & Mn I, Fe I               & 1 \\ 
59 & 5395.75 & 5398.95 & Fe I                     & 1 \\ 
60 & 5399.95 & 5402.05 & Fe I, Cr I               & 1 \\ 
61 & 5403.20 & 5407.00 & Fe I                     & 4 \\ 
62 & 5408.35 & 5411.80 & Cr I, Fe I, Ni I         & 4 \\ 
63 & 5414.45 & 5416.00 & Fe I                     & 1 \\ 
64 & 5416.50 & 5417.60 & Fe I                     & 2 \\ 
65 & 5417.65 & 5419.60 & Ti II                    & 2 \\ 
66 & 5422.90 & 5425.65 & Fe I, Ni I, Fe II        & 4 \\ 
67 & 5427.30 & 5431.00 & Fe I                     & 4 \\ 
68 & 5431.90 & 5435.35 & Fe I, Mn I               & 1 \\ 
69 & 5435.35 & 5437.60 & Ni I, Fe I               & 1 \\ 
70 & 5454.30 & 5456.95 & Fe I                     & 1 \\ 
71 & 5462.00 & 5465.20 & Fe I, Ni I               & 1 \\ 
72 & 5465.20 & 5467.55 & Fe I                     & 1 \\ 
73 & 5469.70 & 5471.75 & Mn I, Fe I               & 1 \\ 
74 & 5475.65 & 5477.45 & Ni I, Fe I               & 1 \\ 
75 & 5482.65 & 5484.25 & Fe I, Co I               & 1 \\ 
76 & 5486.35 & 5488.60 & Fe I                     & 1 \\ 
77 & 5496.80 & 5498.50 & Fe I                     & 1 \\ 
78 & 5500.20 & 5502.45 & Fe I                     & 1 \\ 
79 & 5505.10 & 5507.55 & Fe I, Mn I               & 1 \\ 
80 & 5516.00 & 5517.75 & Mn I, Fe I               & 1 \\ 
81 & 5521.85 & 5522.95 & Fe I                     & 1 \\ 
82 & 5524.70 & 5526.15 & Fe I                     & 1 \\ 
83 & 5531.45 & 5533.80 & Fe I                     & 1 \\ 
84 & 5534.05 & 5536.15 & Fe I, Ba I, Fe II        & 1 \\ 
85 & 5542.30 & 5544.65 & Fe I                     & 3 \\ 
86 & 5545.50 & 5547.85 & Fe I, V I                & 3 \\ 
87 & 5552.90 & 5555.70 & Fe I, Ni I               & 3 \\ 
88 & 5559.40 & 5560.85 & Fe I                     & 3 \\ 
89 & 5562.05 & 5564.50 & Fe I                     & 3 \\ 
90 & 5564.85 & 5566.45 & Fe I                     & 3 \\ 
91 & 5566.80 & 5568.20 & Fe I                     & 3 \\ 
92 & 5568.80 & 5570.30 & Fe I                     & 3 \\ 
93 & 5571.95 & 5574.00 & Fe I                     & 3 \\ 
94 & 5575.40 & 5576.80 & Fe I                     & 3 \\ 
95 & 5577.90 & 5579.25 & Ni I                     & 3 \\ 
96 & 5585.85 & 5587.35 & Fe I                     & 3 \\ 
\enddata
\tablecomments{(1) Indices used in the final analysis; (2) Indices removed 
because their calibrations had $R^{2} <$ 0.9 (see Section \ref{regressions}); 
(3) Indices excluded because they were not present in all 120 solar exposures
(see Section \ref{ewcorr}); (4) Indices eliminated because $\mid\langle 
{\rm EW_{MARVELS}} \rangle - {\rm EW_{FEROS}}\mid\, > \sigma(\langle 
{\rm EW_{MARVELS}} \rangle)$ in the solar spectra (see Section \ref{ewcorr}).} 
\end{deluxetable}


\begin{deluxetable}{crrrrrrrrrrrrrrr}
\tablecolumns{16}
\rotate
\tabletypesize{\scriptsize}
\tablewidth{0pt}
\tablecaption{Calibrations for the Indices.
\label{calibrations_ew}}
\tablehead{ \colhead{Index} & \colhead{$c_{0}$} & \colhead{$c_{1}$} 
& \colhead{$c_{2}$} & \colhead{$c_{3}$} & \colhead{$c_{4}$}
& \colhead{$c_{5}$} & \colhead{$c_{6}$} & \colhead{$c_{7}$}
& \colhead{$c_{8}$} & \colhead{$c_{9}$} & \colhead{$R^{2}$}
& \colhead{$\sigma$} & \colhead{N} & \colhead{EW$_{min}$} 
& \colhead{EW$_{max}$} \\ 
\colhead{} & \colhead{} & \colhead{} & \colhead{} & \colhead{} 
& \colhead{} & \colhead{} & \colhead{} & \colhead{} & \colhead{}
& \colhead{} & \colhead{} & \colhead{(m\AA)} & \colhead{} 
& \colhead{(m\AA)} & \colhead{(m\AA)}} 
\startdata
 1 &  1.75E+03 &  3.28E+02 & -1.04E+00 &  6.82E+02 & -1.17E-02 & -4.34E+01 & -5.57E-02 &  5.88E+01 &  1.06E-04 & -4.39E+01 &  0.895 &  11.55 & 261 &    1.65 &  230.01 \\ 
 2 &  7.65E+02 &  8.84E+02 & -5.20E-01 &  6.13E+02 & -1.97E-02 & -1.36E+02 & -2.72E-02 &  2.39E+01 &  4.35E-05 & -5.49E+01 &  0.926 &  18.39 & 264 &   25.01 &  402.49 \\ 
 3 &  1.04E+02 &  1.37E+02 & -1.23E-01 &  2.43E+02 &  1.94E-02 & -4.35E+01 &   \nodata &  7.23E+00 &  6.21E-06 & -3.02E+01 &  0.935 &   6.06 & 271 &   26.44 &  166.78 \\ 
 4 &  2.02E+03 &  4.08E+02 & -7.82E-01 &  3.48E+02 & -4.32E-02 & -1.31E+01 & -5.91E-02 &   \nodata &  8.09E-05 &   \nodata &  0.974 &   7.48 & 266 &  143.32 &  357.77 \\ 
 5 &  1.68E+03 &  4.54E+02 & -5.41E-01 &  1.37E+02 & -4.75E-02 & -1.74E+01 & -3.68E-02 &  7.28E+00 &  5.32E-05 &  9.16E+00 &  0.980 &   5.56 & 266 &   94.09 &  320.62 \\ 
 6 &  1.24E+03 &  1.42E+02 & -2.62E-01 & -8.39E+01 &  3.16E-02 & -5.26E+01 & -2.75E-02 & -7.08E+00 &  2.87E-05 &  2.75E+01 &  0.955 &   6.90 & 267 &   55.36 &  250.87 \\ 
 7 &  3.60E+03 &  5.65E+02 & -1.15E+00 &  1.65E+02 & -1.52E-02 & -6.65E+01 & -8.45E-02 &  2.62E+01 &  1.20E-04 &  3.87E+01 &  0.964 &  13.58 & 270 &  135.31 &  591.45 \\ 
 8 &  7.51E+03 &  3.36E+02 & -2.94E+00 &  9.48E+02 & -6.03E-02 &  4.85E+01 & -1.47E-01 & -5.02E+01 &  2.85E-04 &   \nodata &  0.986 &  15.23 & 266 &  188.07 &  956.46 \\ 
 9 &  8.65E+03 &  7.99E+02 & -2.87E+00 &  3.53E+02 & -3.08E-02 & -8.81E+01 & -1.49E-01 & -5.27E+01 &  2.83E-04 &  6.37E+01 &  0.975 &  20.20 & 266 &  233.52 & 1006.07 \\ 
10 &  5.71E+03 &  2.62E+02 & -2.22E+00 &  7.34E+02 & -2.53E-02 &  2.64E+01 & -2.11E-01 &   \nodata &  2.52E-04 &  6.17E+01 &  0.985 &  13.55 & 268 &  183.55 &  851.70 \\ 
11 &  9.15E+02 &  7.32E+01 & -2.32E-01 & -4.08E+01 &  1.82E-02 & -2.51E+01 & -3.08E-02 &  4.12E+00 &  3.20E-05 &  2.26E+01 &  0.946 &   4.75 & 266 &   58.48 &  176.57 \\ 
12 &  2.34E+03 &  3.41E+01 & -9.10E-01 &  2.41E+02 &   \nodata &   \nodata & -9.01E-02 & -1.30E+01 &  1.08E-04 &  3.38E+01 &  0.966 &   5.93 & 272 &   50.29 &  242.45 \\ 
13 &  5.47E+03 &  7.34E+01 & -2.10E+00 &  4.87E+02 &   \nodata &   \nodata & -1.15E-01 &   \nodata &  2.15E-04 &  2.24E+01 &  0.976 &   9.78 & 264 &   92.17 &  524.25 \\ 
14 &  3.25E+04 &  5.86E+02 & -1.37E+01 &  4.71E+03 & -2.32E-01 &  2.90E+02 & -7.72E-01 &   \nodata &  1.39E-03 &   \nodata &  0.988 &  46.94 & 263 &  474.31 & 3092.88 \\ 
15 &  4.08E+03 &  8.07E+02 & -1.76E+00 &  8.91E+02 & -1.21E-01 &  4.68E+01 & -1.85E-01 &  7.75E+01 &  2.02E-04 &  2.38E+01 &  0.985 &  13.75 & 260 &  169.39 &  773.83 \\ 
16 &  1.71E+03 &  3.30E+02 & -7.10E-01 &  3.64E+02 & -6.17E-02 &  3.53E+01 & -8.26E-02 &  1.60E+01 &  8.21E-05 &  1.62E+01 &  0.987 &   6.05 & 266 &   69.96 &  360.47 \\ 
17 &  1.90E+03 &  1.86E+02 & -1.37E+00 &  1.37E+03 & -7.87E-02 &  1.13E+02 & -1.50E-01 & -1.23E+01 &  1.51E-04 & -4.58E+01 &  0.991 &   9.96 & 262 &  149.67 &  708.59 \\ 
18 &  2.20E+03 &  3.63E+02 & -2.68E-01 & -4.13E+02 & -1.13E-02 & -1.87E+01 & -2.94E-02 &  4.84E+01 &  2.78E-05 &  6.48E+01 &  0.982 &   8.22 & 271 &  126.15 &  473.52 \\ 
19 &  9.82E+02 &  1.07E+02 & -1.69E-01 & -2.06E+02 & -3.79E-03 & -1.00E+01 & -7.31E-03 &  1.57E+01 &  1.89E-05 &  2.66E+01 &  0.934 &   3.45 & 269 &   32.07 &   99.20 \\ 
20 &  1.16E+03 &  2.46E+02 & -3.95E-01 &  2.04E+02 & -8.85E-03 &   \nodata & -3.79E-02 &  5.90E+01 &  4.20E-05 &   \nodata &  0.968 &   9.63 & 265 &   88.92 &  400.24 \\ 
21 &  7.79E+03 &  5.77E+02 & -2.11E+00 & -2.62E+02 & -7.50E-02 &  3.63E+01 & -3.08E-02 &  1.00E+02 &  1.72E-04 &  5.10E+01 &  0.987 &  13.15 & 269 &  183.32 &  846.77 \\ 
22 &  4.74E+03 &  7.74E+02 & -1.24E+00 & -2.14E+02 & -9.45E-02 &   \nodata &  1.57E-02 &  7.61E+01 &  8.84E-05 &  1.24E+01 &  0.991 &   7.21 & 268 &   94.78 &  541.35 \\ 
23 &  1.93E+03 &  1.49E+02 & -6.57E-01 &  2.18E+02 & -3.25E-02 &  3.88E+01 & -6.54E-02 &  1.00E+01 &  7.16E-05 &  2.11E+01 &  0.988 &   5.30 & 267 &  131.04 &  403.42 \\ 
24 &  3.01E+03 &  5.03E+02 & -8.45E-01 &  9.67E+01 & -4.19E-02 &  1.58E+01 & -4.80E-02 &  6.81E+01 &  8.12E-05 &  1.84E+01 &  0.986 &  10.74 & 271 &  201.06 &  667.33 \\ 
25 &  1.35E+03 &  3.09E+02 & -3.59E-01 &  5.66E+00 & -8.67E-03 & -2.60E+01 & -6.37E-03 &  5.53E+01 &  2.95E-05 &   \nodata &  0.988 &   4.40 & 267 &   47.65 &  265.59 \\ 
26 &  2.97E+03 &  3.25E+02 & -1.21E+00 &  5.37E+02 & -8.44E-02 &  7.94E+01 & -6.54E-02 &  7.01E+01 &  1.15E-04 & -1.43E+01 &  0.990 &   7.13 & 262 &  110.44 &  517.87 \\ 
27 &  3.42E+03 &  3.37E+02 & -1.34E+00 &  6.61E+02 & -7.75E-02 &  9.46E+01 & -5.72E-02 &  7.98E+01 &  1.19E-04 & -3.55E+01 &  0.989 &  10.87 & 264 &  210.44 &  742.70 \\ 
28 &  1.08E+03 &  1.48E+02 & -2.64E-01 & -6.03E+01 & -1.54E-02 &   \nodata &  9.59E-03 &  1.55E+01 &  1.62E-05 &   \nodata &  0.987 &   2.08 & 263 &   33.97 &  135.33 \\ 
29 &  6.57E+03 &  8.00E+02 & -2.22E+00 &  5.23E+02 & -1.02E-01 &  3.75E+01 & -2.14E-01 &  6.90E+01 &  2.45E-04 &  8.84E+01 &  0.991 &  13.90 & 270 &  232.40 & 1148.08 \\ 
30 &  1.69E+03 &  2.85E+02 & -6.89E-01 &  3.40E+02 & -6.82E-02 &  5.45E+01 & -5.33E-02 &  3.26E+01 &  7.00E-05 &   \nodata &  0.987 &   5.38 & 263 &  108.94 &  366.37 \\ 
31 &  1.07E+03 &  3.13E+01 & -3.07E-01 &  6.39E+00 &   \nodata &  5.64E+00 & -1.86E-02 &  1.09E+01 &  3.11E-05 &  1.10E+01 &  0.962 &   3.35 & 269 &   61.56 &  160.74 \\ 
32 &  6.57E+02 &  2.07E+02 & -1.90E-01 & -1.62E+01 & -2.25E-02 & -7.30E+00 &   \nodata &  2.89E+01 &  1.64E-05 &   \nodata &  0.920 &   3.51 & 274 &   11.71 &   88.54 \\ 
33 &  1.18E+03 &  2.70E+02 & -3.39E-01 & -2.06E+01 & -3.22E-02 &   \nodata &   \nodata &  4.33E+01 &  2.79E-05 &   \nodata &  0.971 &   3.67 & 263 &   21.50 &  142.86 \\ 
34 &  6.18E+02 &  2.40E+02 & -1.66E-01 &  4.39E+00 & -3.51E-02 &  6.30E+00 &  1.18E-02 &  3.14E+01 &  7.04E-06 & -8.84E+00 &  0.979 &   2.70 & 262 &   11.12 &  121.67 \\ 
35 &  2.62E+02 &  2.06E+02 &  1.26E-01 & -2.02E+02 & -1.85E-02 &   \nodata &  6.48E-03 &  2.52E+01 & -1.08E-05 &  1.49E+01 &  0.979 &   4.39 & 275 &  102.86 &  269.06 \\ 
36 &  8.56E+02 &  9.15E+01 & -3.10E-01 &  1.85E+02 & -7.42E-03 &  1.45E+01 &   \nodata &  1.52E+01 &  2.02E-05 & -2.22E+01 &  0.976 &   5.83 & 276 &   41.44 &  252.53 \\ 
37 &  1.88E+03 &  2.31E+02 & -1.26E+00 &  1.23E+03 & -8.96E-02 &  1.21E+02 & -1.24E-01 &  2.13E+01 &  1.33E-04 & -4.83E+01 &  0.988 &  11.81 & 273 &  147.63 &  696.29 \\ 
38 &  1.06E+04 &  5.17E+02 & -4.08E+00 &  1.52E+03 & -1.31E-01 &  1.59E+02 & -2.37E-01 &   \nodata &  3.92E-04 &   \nodata &  0.993 &  18.36 & 266 &  369.10 & 1453.86 \\ 
39 &  1.77E+03 &  2.44E+02 & -4.69E-01 & -9.07E+00 & -2.19E-02 &   \nodata &   \nodata &  2.56E+01 &  3.38E-05 &   \nodata &  0.984 &   5.00 & 274 &   68.32 &  295.36 \\ 
40 &  1.73E+02 &  3.03E+01 & -6.04E-02 &  1.46E+01 &  3.33E-03 & -4.77E+00 & -5.47E-03 &  1.41E+01 &  8.31E-06 &   \nodata &  0.901 &   2.90 & 281 &    9.05 &   53.33 \\ 
41 &  8.41E+02 &  8.30E+01 & -1.76E-01 & -1.08E+02 &   \nodata & -1.32E+01 &   \nodata &  1.28E+01 &  1.58E-05 &  1.01E+01 &  0.872 &   3.24 & 282 &   60.27 &  110.59 \\ 
42 &  3.85E+03 &  5.10E+02 & -1.86E+00 &  1.23E+03 & -1.13E-01 &  1.05E+02 & -1.53E-01 &  6.68E+01 &  1.94E-04 & -3.36E+01 &  0.990 &  13.00 & 271 &  208.55 &  892.36 \\ 
43 &  9.45E+02 &  9.38E+01 & -2.98E-01 &  9.32E+01 &   \nodata &   \nodata & -1.65E-02 &  3.09E+01 &  2.66E-05 &   \nodata &  0.960 &   6.21 & 272 &   36.16 &  235.48 \\ 
44 &  2.88E+03 &  3.17E+02 & -1.03E+00 &  2.10E+02 & -3.75E-02 &   \nodata & -5.87E-02 &  2.38E+01 &  1.03E-04 &  1.58E+01 &  0.981 &   5.88 & 268 &   77.20 &  359.05 \\ 
45 &  2.07E+03 &  2.98E+02 & -7.48E-01 &  1.99E+02 & -3.47E-02 &   \nodata & -3.36E-02 &  2.93E+01 &  7.03E-05 &   \nodata &  0.981 &   5.26 & 268 &   49.94 &  302.09 \\ 
46 &  7.37E+02 &  4.55E+01 & -1.28E-01 & -5.95E+01 &  8.18E-03 & -4.45E+00 & -1.22E-02 & -6.81E+00 &  1.16E-05 &  1.51E+01 &  0.981 &   3.39 & 273 &   37.12 &  182.41 \\ 
47 &  1.05E+03 &  1.10E+02 & -1.98E-01 & -1.43E+02 & -7.00E-03 &   \nodata &  2.38E-02 &  1.71E+01 &  5.31E-06 &   \nodata &  0.977 &   3.24 & 270 &   17.62 &  141.32 \\ 
48 &  1.75E+02 & -4.84E+01 &  5.51E-02 & -8.48E+01 &  2.02E-02 &   \nodata &   \nodata &   \nodata & -4.46E-06 &  7.86E+00 &  0.966 &   3.37 & 275 &   64.88 &  158.64 \\ 
49 &  1.37E+03 &  1.83E+02 & -6.10E-01 &  3.99E+02 & -1.86E-02 &  1.80E+01 & -3.15E-02 &  7.72E+00 &  5.66E-05 & -2.22E+01 &  0.980 &   7.33 & 271 &  127.06 &  406.41 \\ 
50 &  8.61E+02 &  1.70E+02 & -4.43E-01 &  3.49E+02 & -3.51E-02 &  3.52E+01 & -2.57E-02 &  1.15E+01 &  4.13E-05 & -2.01E+01 &  0.979 &   5.83 & 269 &   80.91 &  288.34 \\ 
51 &  8.41E+03 &  7.81E+02 & -3.58E+00 &  1.69E+03 & -1.86E-01 &  1.74E+02 & -2.65E-01 &  8.32E+01 &  3.66E-04 &   \nodata &  0.991 &  20.48 & 279 &  274.01 & 1392.36 \\ 
52 &  1.10E+03 &  7.03E+01 & -3.75E-01 &  8.20E+01 &   \nodata &   \nodata & -1.32E-02 &   \nodata &  3.39E-05 &   \nodata &  0.955 &   5.20 & 273 &   28.27 &  180.34 \\ 
53 &  5.06E+02 &  2.42E+02 & -9.25E-02 & -5.18E+00 & -1.94E-02 &   \nodata &   \nodata &  7.31E+01 &  4.05E-06 &   \nodata &  0.978 &   5.19 & 280 &   17.60 &  193.21 \\ 
54 &  1.27E+03 &  2.03E+02 & -8.02E-01 &  7.14E+02 & -4.61E-02 &  4.92E+01 & -3.77E-02 &  1.31E+01 &  7.30E-05 & -5.24E+01 &  0.980 &   8.21 & 268 &  102.63 &  414.96 \\ 
55 &  1.63E+03 &  2.50E+02 & -2.47E-01 & -2.93E+02 & -1.95E-02 &   \nodata &  2.84E-02 &  3.21E+01 &  5.80E-06 &  1.50E+01 &  0.967 &   7.18 & 270 &   26.27 &  237.71 \\ 
56 &  4.56E+03 &  7.69E+02 & -1.31E+00 &  1.32E-01 & -7.90E-02 &   \nodata &   \nodata &  1.04E+02 &  9.97E-05 &   \nodata &  0.979 &  14.32 & 283 &  122.47 &  677.47 \\ 
57 &  3.05E+03 &  1.95E+02 & -9.81E-01 &  1.11E+02 & -5.14E-02 &  5.26E+01 & -1.27E-02 &  1.73E+01 &  7.93E-05 &   \nodata &  0.983 &   6.87 & 269 &   93.58 &  339.92 \\ 
58 &  2.30E+03 &  1.26E+02 & -7.07E-01 &  1.07E+02 & -1.15E-02 &  2.25E+01 &   \nodata &  5.85E+01 &  5.00E-05 & -1.23E+01 &  0.983 &   7.72 & 280 &   23.45 &  304.21 \\ 
59 &  7.81E+03 &  8.83E+02 & -2.82E+00 &  6.88E+02 & -1.23E-01 &  2.92E+01 & -1.46E-01 &  5.02E+01 &  2.73E-04 &  2.62E+01 &  0.987 &  15.22 & 273 &  193.77 &  849.02 \\ 
60 &  2.46E+03 &  3.62E+02 & -8.19E-01 &  1.90E+02 & -2.99E-02 &   \nodata &   \nodata &  4.67E+01 &  6.22E-05 & -2.16E+01 &  0.975 &   9.80 & 284 &   82.59 &  402.08 \\ 
61 &  9.89E+03 &  5.59E+02 & -3.49E+00 &  9.09E+02 & -1.16E-01 &  1.42E+02 & -1.37E-01 &  6.19E+01 &  3.15E-04 &   \nodata &  0.988 &  22.10 & 273 &  320.93 & 1298.99 \\ 
62 &  5.90E+03 &  5.62E+02 & -2.24E+00 &  7.42E+02 & -6.33E-02 &  3.18E+01 & -1.20E-01 &  3.55E+01 &  2.18E-04 &   \nodata &  0.978 &  18.75 & 278 &  220.46 &  984.53 \\ 
63 &  1.52E+03 &  1.53E+02 & -6.41E-01 &  3.37E+02 & -3.86E-02 &  4.74E+01 & -2.00E-02 &   \nodata &  5.46E-05 & -2.07E+01 &  0.964 &   9.44 & 276 &  102.89 &  359.97 \\ 
64 &  1.34E+02 &  2.63E+00 &  3.44E-02 & -6.47E+01 &  7.56E-03 &   \nodata &  1.12E-02 &  9.23E+00 & -8.53E-06 &   \nodata &  0.870 &   4.85 & 277 &   12.81 &   77.37 \\ 
65 &  2.17E+02 &  1.51E+01 &  1.06E-01 & -1.71E+02 &  1.11E-02 &   \nodata &  2.53E-02 &  4.02E+01 & -1.86E-05 &   \nodata &  0.855 &   8.58 & 274 &   47.36 &  155.93 \\ 
66 &  1.96E+03 &  3.94E+02 & -6.06E-01 &  2.19E+02 & -2.02E-02 &   \nodata & -3.18E-02 &  3.11E+01 &  5.30E-05 &   \nodata &  0.969 &  14.72 & 281 &  152.05 &  574.88 \\ 
67 &  7.35E+03 &  8.10E+02 & -2.64E+00 &  6.38E+02 & -9.40E-02 &   \nodata & -9.95E-02 &   \nodata &  2.42E-04 &   \nodata &  0.980 &  18.04 & 277 &  191.54 &  962.50 \\ 
68 &  7.73E+03 &  8.66E+02 & -2.91E+00 &  8.55E+02 & -1.54E-01 &  7.85E+01 & -1.37E-01 &  7.14E+01 &  2.76E-04 &   \nodata &  0.984 &  17.69 & 277 &  177.46 &  845.72 \\ 
69 &  2.80E+03 &  3.30E+02 & -6.29E-01 & -1.94E+02 & -2.18E-02 &   \nodata &  3.22E-02 &  5.09E+01 &  3.18E-05 &   \nodata &  0.984 &   8.41 & 265 &   49.75 &  419.83 \\ 
70 &  4.39E+03 &  3.88E+02 & -1.87E+00 &  8.65E+02 & -9.39E-02 &  8.42E+01 & -1.38E-01 &  2.90E+01 &  1.94E-04 &   \nodata &  0.991 &   9.61 & 269 &  192.69 &  735.89 \\ 
71 &  1.41E+03 &  7.35E+01 & -6.60E-01 &  5.41E+02 &   \nodata &  4.07E+01 & -4.16E-02 &  3.31E+01 &  6.10E-05 & -3.32E+01 &  0.985 &   9.27 & 269 &  129.70 &  536.39 \\ 
72 &  6.93E+02 &  6.16E+01 & -1.99E-01 &  7.77E+01 &   \nodata &  1.21E+01 & -1.33E-02 &  2.16E+01 &  1.74E-05 &   \nodata &  0.976 &   5.31 & 274 &   47.53 &  236.68 \\ 
73 &  2.68E+03 &  5.54E+02 & -8.35E-01 &  6.33E+01 & -7.21E-02 &   \nodata & -1.09E-02 &  6.26E+01 &  6.71E-05 &   \nodata &  0.987 &   5.89 & 270 &   20.75 &  300.08 \\ 
74 &  1.19E+03 &  1.13E+02 & -4.96E-01 &  4.82E+02 & -1.03E-02 &  4.23E+01 & -7.88E-02 &  3.26E+01 &  5.82E-05 &   \nodata &  0.990 &   8.03 & 276 &  202.99 &  604.88 \\ 
75 &  1.24E+03 &  2.90E+02 & -1.33E-01 & -2.23E+02 & -1.82E-02 & -1.47E+01 &  1.99E-02 &  4.56E+01 & -3.53E-06 &  1.16E+01 &  0.982 &   5.58 & 280 &   17.49 &  237.20 \\ 
76 &  2.32E+03 &  5.40E+02 & -6.80E-01 &  7.71E+01 & -4.92E-02 & -1.17E+01 & -1.32E-02 &  7.14E+01 &  5.41E-05 &   \nodata &  0.990 &   6.64 & 282 &   57.41 &  405.60 \\ 
77 &  2.91E+03 &  3.00E+02 & -9.33E-01 &  8.25E+01 & -2.74E-02 & -9.92E+00 & -5.32E-02 &  1.81E+01 &  9.31E-05 &  2.71E+01 &  0.983 &   5.63 & 271 &   80.82 &  346.14 \\ 
78 &  1.88E+03 &  3.87E+02 & -6.33E-01 &  1.53E+02 & -3.50E-02 & -1.61E+01 & -4.72E-02 &  3.82E+01 &  6.64E-05 &  1.30E+01 &  0.972 &   6.41 & 279 &   87.22 &  317.56 \\ 
79 &  4.64E+03 &  6.55E+02 & -1.63E+00 &  3.08E+02 & -7.55E-02 & -1.35E+01 & -5.24E-02 &  3.51E+01 &  1.48E-04 &   \nodata &  0.983 &   8.92 & 274 &   90.97 &  506.80 \\ 
80 &  2.28E+03 &  6.35E+02 & -6.00E-01 & -9.05E+01 & -6.72E-02 & -2.44E+01 &  1.44E-02 &  6.82E+01 &  3.91E-05 &   \nodata &  0.984 &   5.91 & 274 &    4.81 &  262.53 \\ 
81 & -1.78E+01 &  1.25E+01 &  4.93E-02 & -3.53E+00 &  5.45E-03 &   \nodata &   \nodata &   \nodata & -6.20E-06 &   \nodata &  0.944 &   2.97 & 275 &   10.24 &   73.26 \\ 
82 &  2.33E+02 &  1.61E+02 & -3.68E-02 &  3.14E+01 &   \nodata & -1.74E+01 & -7.89E-03 &  2.53E+01 &  3.65E-06 &   \nodata &  0.973 &   3.82 & 273 &   23.46 &  145.70 \\ 
83 &  1.68E+03 &  5.67E+02 & -4.07E-01 & -8.50E+01 & -5.96E-02 & -2.29E+01 &  1.31E-02 &  5.50E+01 &  2.38E-05 &   \nodata &  0.958 &   8.34 & 282 &   17.03 &  237.91 \\ 
84 &  4.80E+02 &  3.10E+02 & -2.64E-02 & -3.56E+01 & -9.49E-03 & -2.56E+01 &   \nodata &  2.61E+01 &   \nodata &   \nodata &  0.969 &   7.01 & 287 &   70.26 &  272.69 \\ 
85 &  9.04E+02 &  3.19E+02 & -3.12E-01 &  1.80E+02 &   \nodata & -4.04E+01 &   \nodata &  1.73E+01 &  2.16E-05 & -2.32E+01 &  0.943 &  10.38 & 289 &   50.34 &  288.82 \\ 
86 &  1.60E+03 &  4.26E+02 & -4.31E-01 & -2.13E+01 & -2.43E-02 & -3.37E+01 &   \nodata &  5.65E+01 &  3.27E-05 &   \nodata &  0.949 &   9.23 & 288 &   27.81 &  269.08 \\ 
87 &  8.09E+02 &  4.29E+02 & -1.23E-01 &  5.13E+00 & -2.48E-02 & -2.56E+01 &  2.35E-02 &  5.00E+01 & -5.56E-06 & -1.81E+01 &  0.977 &   7.58 & 279 &   59.74 &  312.23 \\ 
88 &  2.96E+02 &  1.37E+02 & -1.74E-03 & -6.35E+01 &   \nodata & -1.56E+01 &  9.61E-03 &  1.55E+01 & -5.55E-06 &   \nodata &  0.956 &   4.04 & 290 &   13.13 &  117.75 \\ 
89 &  1.43E+03 &  3.10E+02 & -2.88E-01 & -6.82E+01 & -1.04E-02 & -1.97E+01 &  1.02E-02 &  2.95E+01 &  1.36E-05 &   \nodata &  0.986 &   5.70 & 275 &   62.96 &  297.99 \\ 
90 &  1.17E+03 &  3.31E+02 & -3.68E-01 &  1.11E+02 & -3.58E-02 &   \nodata & -1.81E-02 &  3.37E+01 &  3.18E-05 &   \nodata &  0.989 &   4.28 & 275 &   48.27 &  262.34 \\ 
91 &  9.60E+02 &  1.77E+02 & -2.73E-01 &  4.37E+01 & -6.48E-03 & -1.01E+01 &   \nodata &  2.81E+01 &  1.90E-05 & -6.29E+00 &  0.989 &   3.00 & 266 &   22.05 &  157.34 \\ 
92 &  1.40E+03 &  3.18E+02 & -6.53E-01 &  4.47E+02 & -6.81E-02 &  4.72E+01 & -5.77E-02 &  1.23E+01 &  6.69E-05 & -8.61E+00 &  0.991 &   4.93 & 263 &   83.85 &  347.46 \\ 
93 &  1.85E+03 &  3.10E+02 & -8.84E-01 &  6.41E+02 & -6.29E-02 &  5.64E+01 & -5.19E-02 &  9.02E+00 &  8.04E-05 & -3.39E+01 &  0.990 &   7.30 & 271 &  113.74 &  474.63 \\ 
94 &  1.11E+03 &  2.05E+02 & -5.11E-01 &  3.19E+02 & -3.74E-02 &  2.15E+01 & -3.49E-02 &  3.73E+00 &  5.06E-05 & -1.17E+01 &  0.986 &   3.77 & 265 &   62.06 &  265.81 \\ 
95 &  4.04E+02 &  1.00E+02 &  4.59E-02 & -1.45E+02 &  1.62E-02 & -2.85E+01 &  2.16E-02 &  2.72E+01 & -1.56E-05 &   \nodata &  0.956 &   4.74 & 270 &   12.75 &  129.88 \\ 
96 &  1.36E+03 & -5.09E+01 & -9.19E-01 &  9.02E+02 & -2.51E-02 &  8.31E+01 & -9.26E-02 & -2.45E+01 &  9.79E-05 & -3.40E+01 &  0.988 &   8.72 & 266 &  113.80 &  514.74 \\ 
\enddata
\end{deluxetable}


\begin{deluxetable}{lrrrc}
\tablecolumns{5}
\tablewidth{0pt}
\tablecaption{Comparison Between Atmospheric Parameters from Spectral
Indices and High-Resolution Analyses.
\label{diff_param}}
\tablehead{ \colhead{Sample} & \colhead{$\Delta$\teff} & 
\colhead{$\Delta$[Fe/H]} & \colhead{$\Delta$\logg} & \colhead{Number of} \\ 
\colhead{} & \colhead{(K)} & \colhead{} & \colhead{} & \colhead{Stars/Spectra}} 
\startdata
\sidehead{\textit{Complete Set of 92 Indices}}
Calibration &     1 $\pm$ 78  & 0.01 $\pm$ 0.06 &    0.00 $\pm$ 0.15 & 309 \\
MARVELS     & $-$97 $\pm$ 110 & 0.00 $\pm$ 0.07 & $-$0.30 $\pm$ 0.17 &  30 \\
ELODIE      &    15 $\pm$ 125 & 0.06 $\pm$ 0.10 &    0.07 $\pm$ 0.29 & 219 \\
\sidehead{\textit{Restricted Set of 64 Indices}}
Calibration &  $-$3 $\pm$ 80  & 0.01 $\pm$ 0.06 &    0.00 $\pm$ 0.15 & 309 \\
MARVELS     & $-$28 $\pm$ 81  & 0.02 $\pm$ 0.05 & $-$0.07 $\pm$ 0.15 &  30 \\
\enddata
\end{deluxetable}


\end{document}